\documentclass[aps, prd,nofootinbib,showpacs,superscripttaddress,preprint]{revtex4-2}
\usepackage{amssymb}
\usepackage{amsmath}
\usepackage{graphicx}
\usepackage[colorlinks=true, citecolor=blue, urlcolor = blue, linkcolor= red, bookmarks=true]{hyperref}
\usepackage{orcidlink}

\ExplSyntaxOn
\NewDocumentCommand{\RN}{m}
 {
  \textup{ \int_to_Roman:n { #1 } }
 }
\ExplSyntaxOff

\newcommand{\beq}{\begin{equation}} \newcommand{\eeq}{\end{equation}}
\newcommand{\bea}{\begin{eqnarray}} \newcommand{\eea}{\end{eqnarray}}

\newcommand{\lsim}{\mathrel{\hbox{\rlap{\lower.55ex\hbox{$\sim$}} \kern-.3em \raise.4ex \hbox{$<$}}}}
\newcommand{\gsim}{\mathrel{\hbox{\rlap{\lower.55ex\hbox{$\sim$}} \kern-.3em \raise.4ex \hbox{$>$}}}}

\def\lsim{\mathrel{\raise.3ex\hbox{$<$\kern-.75em\lower1ex\hbox{$\sim$}}}}
\def\gsim{\mathrel{\raise.3ex\hbox{$>$\kern-.75em\lower1ex\hbox{$\sim$}}}}

\newcommand{\be}{\begin{eqnarray}}
\newcommand{\ee}{\end{eqnarray}}

\newcommand{\benum}{\begin{enumerate}}
\newcommand{\eenum}{\end{enumerate}}
\newcommand{\bi}{\begin{itemize}}
\newcommand{\ei}{\end{itemize}}

\begin{document}

\title{Naturalness in Soft Leptogenesis and Gravitino \\ Mass Bound due to Primordial Black Holes }

\author{Suhail Khan$^{a}$\orcidlink{0009-0007-4941-0069}}
\thanks{Email: \href{mailto:suhail@ctp-jamia.res.in} {suhail@ctp-jamia.res.in}}

\author{Rathin Adhikari$^{a}$\orcidlink{0000-0002-8764-9587}}
\thanks{Email: \href{mailto:rathin@ctp-jamia.res.in}{rathin@ctp-jamia.res.in}} 

\affiliation{$^a$Centre for Theoretical Physics, Jamia Millia Islamia (Central University), \\ New Delhi-110025}

\begin{abstract}
If sneutrinos are produced through primordial black hole (PBH) evaporation, then some interesting features of soft leptogenesis in the Minimal Supersymmetric Standard Model with heavy right-handed neutrinos, are found. The required baryonic asymmetry could be possible from the decays of sneutrinos, for the soft SUSY breaking trilinear $A$ and bilinear $B$ parameters around the electroweak scale. The resonance condition in soft leptogenesis is not required. The allowed regions of different relevant parameters are discussed in detail. Using experimental constraints from collider searches on heavy leptons, the lower bound on the right-handed neutrino mass is found to be around 
$300$ GeV with Yukawa coupling lesser than about $0.4$.
 The allowed region of the typical mass scale of some supersymmetric particles and the $|A|$ parameter is also shown from the experimental constraint on the branching ratio for $\mu \rightarrow e \gamma$ in  MEG \RN{2} search. 
Depending on PBH mass,  bounds on the mass of gravitinos,
produced from PBH evaporation, is discussed and gravitino mass around the electroweak scale is found to be possible only for unstable gravitino.

\end{abstract} 
\maketitle
\newpage
\section{Introduction}
In our observed universe, the number density of the baryons ${n_B}$ is found to be more than the number density of anti-baryons $n_{\bar B}$  which is characterized by baryonic asymmetry ~\cite{asym,Cyb},
\be\label{first}
Y_{\Delta{B}} \equiv \frac{n_B - n_{\bar B}}{s} \approx 8.6\pm 0.1\times 10^{-11}
\ee
where $s$ is the entropy density of the universe.
 In general, the mechanism of baryogenesis could explain the above asymmetry by considering Sakharov's three conditions. These are as follows: $(1)$ Baryon number or Lepton number violating interaction must be present in the Lagrangian $(2)$ There must be $C$ and $CP$ violating physical processes $(3)$ Such physical processes must be in out of thermal equilibrium \cite{Sakharov}. Moreover, one also requires an interference term involving the amplitudes of tree-level and higher-order Feynman diagrams corresponding to lepton/baryon number violating physical processes which will give $CP$ asymmetry. The amplitude of the higher-order diagram should have the absorptive part of the loop integral. Furthermore, in the higher order diagram, total $B$ or $L$ violation through couplings (corresponding to baryogenesis or leptogenesis) should be present with final states on the right of the "cut"(for which on-shell condition is satisfied on the internal line)~\cite{Adhikari:2001yr}. 
There are various  works like GUT baryogenesis~\cite{Harvey:1981yk,Weinberg:1979bt,Nanopoulos:1979gx,Ignatiev:1978uf,Yoshimura:1979gy,Riotto:1998bt}, electroweak baryogenesis~\cite{Klinkhamer:1984di,Kuzmin:1985mm,Arnold:1987mh,Arnold:1987zg,Khlebnikov:1988sr,Kajantie:1996mn,Riotto:1999yt,Cline:2006ts,Sami:2021ufn}, leptogenesis~\cite{Fukugita:1986hr,Buchmuller:2004nz,Buchmuller:2005eh,Pilaftsis:2005rv}, where Sakharov's conditions are useful. There are other ways to generate asymmetry like the Affleck-Dine mechanism~\cite{Affleck:1984fy,Dine:2003ax} in which flat directions of the scalar potential of supersymmetry have been considered.  After the discovery of the mass of Higgs boson  around $125$ GeV at LHC \cite{CMS:2020xrn}
electroweak baryogenesis is found to be not so much preferred \cite{Cline:2017jvp}
because of the difficulties in obtaining appropriate first-order phase transition of the universe at the electroweak scale. 

There are interesting works on GUT scale baryogenesis ~\cite{Harvey:1981yk,Weinberg:1979bt,Nanopoulos:1979gx,Ignatiev:1978uf,Yoshimura:1979gy,Riotto:1998bt}, in which, a highly massive particle around $10^{16}$ GeV near the GUT scale was considered. However, the energy scale of inflation may be somewhat lower, because of the constraint on tensor to scalar ratio from the Planck and BICEP$2$ experimental data \cite{Tristram:2020wbi} and because of that baryogenesis scale may be considered to be lower also. On the other hand, leptogenesis seems to be another attractive mechanism for the generation of baryonic asymmetry in which lepton number-violating interactions are required.  Such lepton number violation could lead to baryon number violation in the presence of sphaleron transitions. There are several works in leptogenesis in which leptonic asymmetry is generated due to the decays of heavy right-handed neutrinos \cite{Fukugita:1986hr,Buchmuller:2004nz,Buchmuller:2005eh}. But to avoid the gravitino overproduction affecting the successful predictions of Big Bang Nucleosynthesis (BBN), the reheat temperature $T_{reheat}$  of the universe, should be lower than $10^6 - 10^9$ GeV for gravitino mass $0.2-8 $ TeV \cite{
JCAP.0910.021,PhysRevD.78.065011,PhysRevD.75.075011,PhysRevD.79.103534,ModPhysLettA.23.427,PhysLettB.648.224,PhysRevD.75.023509,PhysRevD.71.083502,JHEP.9911.036,PhysRevD.67.103521,PhysRevD.67.103521,NuclPhysB.606.518,ProgTheorPhys.93.879,PhysLettB.303.289,PhysAtomNucl.57.1393,PhysLettB.189.23,NuclPhysB.259.175,NuclPhysB.259.175,PhysLettB.158.463,PhysLettB.145.181,PhysLettB.138.265,PhysRevD.79.103534,Davidson:2002qv}. This requires that the  mass of the lightest right-handed neutrinos
(RHN) to be less than $T_{reheat}$.

However, for successful leptogenesis with hierarchical heavy RHN, the mass of RHN in general, is required to be greater than about $10^9$ GeV \cite{Davidson:2002qv}.
So there is an apparent conflict with the requirement of RHN masses to avoid gravitino overproduction and to get successful leptogenesis. However, in resonant leptogenesis, considering almost degenerate two RHN masses, one may get successful leptogenesis with light RHN mass \cite{Pilaftsis:2005rv}. Without considering such almost degeneracy, this conflict on the requirement of heavy RHNs mass may be resolved in soft leptogenesis in supersymmetry  \cite{D'Ambrosio:2003wy,Chung:2003fi,Grossman:2003jv,Fong:2011yx,Fong:2010zu,Grossman:2004dz,Fong:2009iu,Garbrecht:2013iga,Fong:2010bv,Kashti:2004vj,Abdallah:2020,Medina:2006,Adhikari:2015ysa}.
In this scenario, one considers Type I seesaw supersymmetric leptogenesis in which decays of sneutrinos could produce lepton number asymmetry. The Type I seesaw mechanism also could explain the origin of very light active neutrino masses.  There are various soft supersymmetry breaking parameters like trilinear $A$, and bi-linear $B$ parameters and also there are Yukawa couplings $Y$ among RHNs, lepton, up type Higgs superfields and all these could provide the required amount of $CP$ violation. There is mixing between heavy sneutrinos and anti-sneutrinos in the presence of soft supersymmetry breaking terms and there is a small mass splitting between two sneutrino mass eigenstates due to the above mixing. The mass splitting is of the order of the $B$ parameter. It is possible to get successful leptogenesis for right-handed sneutrino mass $\lesssim 10^7$ GeV when $B$ is of the order of the decay width of sneutrinos.  Thus the problem of gravitino overproduction could be to some extent evaded for very small $B$ parameters.  However, $B$ is naturally expected to be not too small but around the electroweak scale in general. 

There are recent developments in the work on baryogenesis where decaying heavy particles are produced through PBH evaporation. 
The primordial black holes (PBH) \cite{carr:2021,Escriva:2022duf,Khlopov:2008qy} due to quantum effect, undergo evaporation through Hawking radiation \cite{Hawking:1974,Hawking:1975}. PBH with mass greater than $10^{37}$ GeV may exist even today, which may play the role of dark matter  \cite{Kuhnel:2016,Frampton:2016,García:2017,Anne:2021,carr:2022,ShamsEsHaghi:2022azq,Belotsky:2014kca,Gangopadhyay:2021kmf}. However, PBH mass lesser than that could evaporate before BBN and could have various interesting cosmological consequences \cite{Carr:2010}.  Such PBH  may lead to the production of heavy particles in the early universe and the decay of such particles could play the role of baryogenesis or leptogenesis to create observed baryonic asymmetry of the universe. There are several earlier works in which PBH evaporation mechanisms have been considered in the context of baryogenesis \cite{Nolan:2022,Tomohiro:2104,Hamada;2017,Hooper:2021,Luca:2021,Ghosal:2021,Pinetti:2022,Stefano:2019,Fujita:2014hha,Gehrman:2022imk} and leptogenesis \cite{Gonzalez:2021,Nicolás:2022,Borah:2021}.

We consider the heavy particle $X$  having a lepton/baryon number violating decay modes could be produced due to PBH evaporation. $T_{\mbox{evap}}$ is  the temperature  after PBH complete evaporation. There are two cases: one where $T_{\mbox{evap}} \gtrsim M_{X}$ and the other is  $T_{\mbox{evap}}< M_{X}$ where $M_{X}$ is the mass of the decaying particle $X$ with baryon/lepton number violating decay modes.  In case, the first condition is satisfied then $X$ particles produced through evaporation will be in thermal equilibrium with other lighter particles. However, in case the second condition is satisfied then $X$ particles produced through evaporation will be out of thermal equilibrium with other lighter particles. For initial PBH mass $M_{BH}$, there is corresponding initial Hawking temperature $T_{BH}$ \cite{temphawk}. There are two scenarios. In one case  $T_{BH}\gtrsim M_{X}$ and in the other case $T_{BH} \lesssim M_{X}$. The expression of baryonic asymmetry is different depending on these conditions. This is because the number density of the heavy particle $X$ produced through PBH evaporation depends differently on $M_{BH}$ and $M_X$ in these two cases. Combining all these, we have classified the following four different cases: (\RN{1}) $T_{\mbox{evap}}\gtrsim M_{X}$ and  $T_{BH} > M_{X}$, (\RN{2})  $T_{\mbox{evap}}< M_{X}$ and  $T_{BH}>M_{X}$, (\RN{3})  $T_{\mbox{evap}}< M_{X}$ and  $T_{BH} \lesssim M_{X}$, (\RN{4}) $T_{\mbox{evap}}\gtrsim M_{X}$ and  $T_{BH} \lesssim M_{X}$. We have shown that for case (\RN{4}) there is no allowed region in $M_X$ versus $M_{BH}$ plane. For the other three cases, one could get, in principle, some leptonic/baryonic asymmetry. However, as shown later, considering $X$ as sneutrino if we consider the generation of baryonic asymmetry through sneutrino decays in Minimal Supersymmetric Standard Model with heavy right-handed neutrinos and consider soft SUSY breaking parameters $A$ and $B$ to be around electro-weak scale, then only case (\RN{2}) is found to be appropriate. 

In section  \ref{sec:bh}, there is a brief discussion on how PBH is formed in the early universe and what are their lifetime and mass due to constraints from the BBN epoch. There is a discussion on the production of heavy particles due to the evaporation of PBHs, the number density of such heavy particles, and the leptonic asymmetry due to the lepton number violating decay modes of the heavy particles and their antiparticles. There is a discussion on the allowed region of $M_X$ versus $M_{BH}$ for four different cases as mentioned earlier.
In section \ref{sec:baryon}, we discuss the interaction of the heavy RHNs, lepton, and Higgs superfields in the mass eigenstate basis in the Minimal Supersymmetric Standard Model (MSSM) and also discuss non-thermal $CP$ asymmetry generated in the decays of sneutrinos in the early universe. Based on $T_{BH} > M_X$ and $T_{BH} \lesssim M_X $ two expressions of leptonic asymmetry in MSSM are discussed.
Based on required baryonic asymmetry as observed and the CMS, ATLAS constraints on right-handed neutrino masses and MEG II experimental constraints on $\mu \rightarrow e \gamma$, the allowed regions of various SUSY parameters and $M_{BH}$ have been shown.
In section \ref{natural}, we have discussed the allowed region of mass of gravitino versus PBH mass for unstable and stable gravitino produced from PBH.
Concluding remarks are given in section \ref{sec:summary}.

\section{Heavy particles production from PBH and leptonic asymmetry}
\label{sec:bh}
The idea of the formation of a black hole in the early universe was initially put forward by Zel'dovich and Novikov~\cite{Zeldovich:1966vw} and subsequently, the formation of a primordial black hole  was considered for the post-inflationary period with density fluctuation~\cite{Carrhawk1974,Carr:1975qj}. Also, the formation was proposed in the context of hybrid inflation~\cite{GarciaBellido:1996qt}.  

When the radiation pressure is insufficient to withstand the gravitational collapse in dense regions during the radiation-dominant epoch, PBHs are created. Then the mass of PBH at the time of formation is proportional to its horizon mass\cite{Carr:2010,Zel’dovich:1966,Hawking:1971,Carr:1974,Polnarev:1979} and is given by,

\begin{eqnarray}\label{hmass}
 M_{BH} =\gamma M_H = \gamma\frac{4\pi}{3} \rho H_f^{-3} =\gamma \frac{4\pi}{3} \frac{3 H_f^2}{8 \pi G} H_f^{-3}= \gamma \frac{1}{2GH_f}=\frac{ \gamma M_{Pl}^2}{2 H_f}.
\end{eqnarray}
Here, $M_{BH}$ is the black hole mass, and $M_{Pl} = 1.22 \times 10^{19}$ GeV is the Planck mass and $\gamma$ is a numerical parameter that is affected by the mechanics of gravitational collapse.
This $\gamma$ has been estimated to be around $(1/\sqrt{3})^3 \approx 0.2$  during the radiation period \cite{Carr:1974,Nadezhin1978,Shibata1999,Musco,Harada2013,Nakama}.  $\rho$ is the universe's average total energy density, $H_f$ is the Hubble parameter at the time of formation and $1/H_f$ is the measure of the size of the horizon at that time. Planck collaboration \cite{PAR} has put an upper bound on the Hubble parameter during inflation at 95\% confidence level as $H_{inf}  < 3.7 \times 10^{-5} M_{P} \sim  10^{14}$ GeV. As $H_f < H_{inf}$, using this condition in Eq.~(\ref{hmass}), one gets the lower bound on PBH mass at formation as

\be\label{T}
M_{BH}\gtrsim \frac{ \gamma M_{Pl}^2}{2 H_f}
\gtrsim 1.49 \times 10^{23} \rm GeV.
\ee
\\ 

The Hawking temperature of non-rotating, electrically neutral (Schwarzschild) PBH is
\begin{eqnarray}
\label{temp}
{ T}_{BH} = \frac{M^2_{Pl}}{8\pi{M}_{BH}} \simeq 5.92\times 10^{36} \bigg(\frac {  \rm 1 \; GeV}{{M}_{BH}} \bigg)  \rm GeV ,
\end{eqnarray}
Loss of PBH mass will occur due to the radiation of particles through Hawking evaporation.  Using the time scale $\tau_{\rm evap}$ for the full evaporation  of PBH  \cite{Hawking:1974sw} and the Friedman equation
one may write the energy density of PBH as
\be
\label{TRH1}
\rho_{BH} = \frac{ 3 M_{Pl}^2 \;H(\tau_{\rm evap})^2 }{ 8\pi }    =  
\frac{ M_{Pl}^2 }{ 6\pi \tau_{\rm evap}^2    }  =5.12 \times 10^{-11} g^2 \;\Lambda^2\;\bigg(\frac{ M_{Pl}^{10}}{ \; {M_{BH}}^6}\bigg) \rm.
\ee
where $g$ represents an effective number of the particle's degrees of freedom. In Minimal Supersymmetric Standard Model (MSSM)  $g \simeq 316$~\cite{Keith2016} for the full particle content of the model and $\Lambda $ is the relevant blackhole greybody factor $ \approx 3.8$ \cite{Hooper:2021}. 

During the expansion of the universe, the ratio of the energy density of PBH to the energy density of radiation is $\rho_{BH}/\rho_{rad} \propto a $ which grows with the expansion of the universe ~\cite{Morris,Krnjaic:2019,March-Russell:2018}. Therefore, even if PBHs are initially subdominant, the slower rate of dilution allows them to eventually dominate the energy density of the Universe.
In the early universe, PBHs could dominate the total energy density before they complete their evaporation if their initial energy density, \( \rho_{\text{BH},i} \), relative to the initial radiation energy density, \( \rho_{\text{R},i} \), satisfies the following condition \cite{Hooper:2021}  
\[
\frac{\rho_{\text{BH},i}}{\rho_{\text{R},i}} \gtrsim 4 \times 10^{-9} \left( \frac{10^{10}\,\text{GeV}}{T_i} \right) \left( \frac{5.62 \times 10^{29}\,\text{GeV}}{M_{\text{BH},i}} \right)^{3/2}
\]
where, \( T_i \) denotes the initial temperature of the radiation, and \( M_{\text{BH},i} \) represents the initial mass of the black holes. 

\begin{figure}[htb]
\includegraphics[scale=.75]{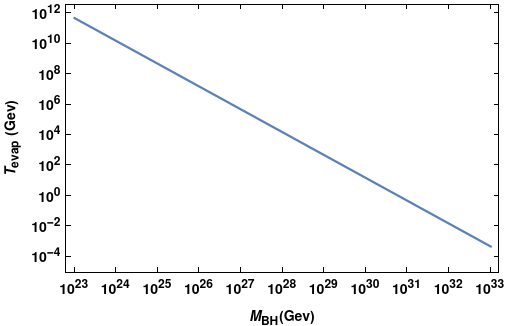}
\caption{$T_{\mbox{evap}}$ versus $M_{BH}$ plot using Eq.~(\ref{tevap}).} 
\label{mbhevap}
\end{figure}

Due to the evaporation of black holes, there will be reheating of the universe to a temperature, $T_{\mbox{evap}}$. As the particle formation due to PBH evaporation is  almost instantaneous \cite{Baumann:2007yr},  all black hole energy density is transformed to radiation and  
$
\rho_{BH} \approx
103.96 \, T_{\mbox{evap}}^4 \rm GeV^4.$
Then $T_{\mbox{evap}}$ can be written as

\be \label{tevap}
T_{\mbox{evap}}= \frac{{g^{1/4}} {\Lambda^{1/2}} M_{Pl}^{5/2}}{67.67   \pi ^{5/4} M_{BH}^{3/2}} \approx 1.51\times 10^{46} \bigg(\frac { \rm 1.GeV}{M_{BH}} \bigg)^{3/2} \rm GeV
\ee
This $T_{\mbox{evap}}$ could be related to the reheating temperature of the universe under certain conditions on the type of inflaton potential and the PBH energy density \cite{RiajulHaque:2023cqe}. 
The fraction $\beta_{PBH}$ of PBH energy density to the total energy density $\rho_{\text{tot}}$ which is the sum of the inflaton energy density and the radiation energy density,  is written as

\be
\beta_{PBH} = \frac{\rho_{\text{BH}}(t_{\text{in}})}{\rho_{\text{tot}}(t_{\text{in}})}
\ee
where $t_{\text{in}}$ correspond to the time of formation of PBH. For certain critical values of $\beta_{PBH}$, the PBH energy density dominates the energy budget of the universe over the inflaton field. Then all the entropies generated due to the decays of PBH would be transferred to the thermal bath.  For a quartic inflaton potential and for PBH mass about $10^{24}$ GeV such a critical value of $\beta_{PBH}$ is shown in Fig 7 (a) of \cite{RiajulHaque:2023cqe} to be about $10^{-6}$ and this decreases for further higher initial PBH masses.  With $\beta_{PBH} > \beta_{PBH}^{critical}$ the reheating temperature $T_R$ is independent of $\beta_{PBH}$ and  $T_R \propto M_{BH}^{-3/2} $.  Under such consideration  $T_{\mbox{evap}}$ may be approximately treated as $T_R$. In this work, we will consider  that the energy density of the universe is dominated by the PBH and its evaporation leads to a radiation dominated universe and evaporation temperature is the reheating temperature of the universe. Due to a slower rate of dilution ($\propto a^{-3}$), PBH dominates over inflation and this results in universal reheating temperature as about $T_{\mbox{evap}}$ and erases initial information about inflation \cite{RiajulHaque:2023cqe}. In such a scenario, following Eq.~(\ref{tevap}) the variation of PBH mass $M_{BH}$ with $T_{evap}$ is shown in Figure \ref{mbhevap}.

The black hole should have evaporated before nucleosynthesis and the universe is required to be radiation-dominated at nucleosynthesis for reproducing the success of Standard Big-bag cosmology \cite{Baumann:2007yr}. So
\be
T_{\mbox{evap}} > T_{\mbox{BBN}}
\ee
Considering $T_{\rm BBN} \sim 1 $ MeV and using Eq.~(\ref{tevap}), one obtains the upper bound on primordial black hole mass as
\be\label{up}
 M_{BH} \lesssim 6.1 \times 10^{32} \mbox{GeV} 
\ee
for PBH producing heavy particles through evaporation.
However, it could be possible for PBH to have a mass much higher than $10^{32}$ GeV, for which the evaporation rate of PBH is too slow and their lifetime is nearer to the age of the universe, and due to that  BBN will not be disturbed by the presence of such PBH. In this work, we are not interested in such PBH with a very low evaporation rate.

Particles of different masses will be produced through Hawking evaporation of PBHs. However, we are interested in the production of scalar $X$ particles (the sneutrinos which are super-partners of heavy right-handed neutrinos in MSSM in our work) which could be radiated through evaporation. It will be demonstrated that leptonic asymmetry may be formed from the decay of such particles. Taking into account the mass of $X$ particle $M_X < T_{BH}$ and integrating over the Bose-Einstein distribution, the number of $X$ particles produced by a single black hole evaporation, is
\be
\label{n<x}
N_{<X}= \frac{4 \pi \; {f_X} \; {M_{BH}}^2} {3 M_{Pl}^2}
\ee
 where $f_X$ is given by 
\be
f_X \sim g_X/g
\ee
and $g_X $ is the number of degrees of freedom of particle $X$ \cite{Baumann:2007yr,MacGibbon:1990,MacGibbon:1991} and $g$ is the total number of degrees of freedom as mentioned earlier. The lepton asymmetry parameter is defined as 
\be
\label{yl}
Y_{\Delta{L}}=\frac{n_L - {n}_{\bar{L}}}{s}
\ee
where $n_L$ and ${n}_{\bar{L}}$ is the number density of leptons and anti-leptons respectively of the universe.  $\eta_X$ is $CP$ asymmetry parameter defined as 
\begin{equation}
\label{etax}
\eta_{X} = \sum_f L_f \frac{ \Gamma_{X}({{X}} \rightarrow f )-\Gamma_{X}(\tilde{{X}}\rightarrow \bar{f} )}{\Gamma_{X}},
\end{equation}
where $L_f$  is a change in lepton number produced through the lepton number violating decay modes of $X$ particles with $f$ as final states and $\Gamma_X$ is the decay width of particle $X$. The sum runs over all lepton numbers violating final states. 
 After using Eq.~(\ref{etax}), in Eq. (\ref{yl}), the  lepton asymmetry $Y_{\Delta{L}}$  can be written as,
\begin{eqnarray}
\label{BHdom}
Y_{\Delta{L}} &=& \frac{n_{BH}}{s} \,  \eta_X N_{<X} =\frac{\eta_X  f_X {g}^{1/4}  }{32 \sqrt{2} \sqrt[4]{5 \pi} }  \left( \frac{ M_{Pl}}{M_{BH}}\right)^{1/2}\approx 10^{-2}   f_X \;{g}^{1/4}\;   \eta_{X }  \left( \frac{ M_{Pl}}{M_{BH}}\right)^{1/2}
\ee
where $n_{BH}={\rho_{BH} }/{{M_{BH}}}$ and   entropy density $s$ of the universe  at the end of evaporation is given by
$$
s = \bigg(\frac{2}{45} \pi ^2 g \; T_{\mbox{evap}}^3\bigg) \rm GeV^3.
\ 
$$.

For $M_X \gtrsim T_{BH}   $, the number of $X$ particles produced by a single black hole evaporation, is  
\be\label{n>x}
 N_{\gtrsim X} = \frac{f_X \;{ M_{Pl}}^2}{48 \pi \; {M_X}^2} \; .
\ee 
The lepton asymmetry generated through lepton number violating decay modes is given by
\be\label{ydl}
Y_{\Delta{L}}=\frac{n_{BH}}{s} \,  \eta_X N_{>X} \approx{1.7 \times 10^{-5}\; f_X \; g^{1/4}}\;\eta_X  
\left( \frac{ M_{Pl}^9}{M_{BH}^5 M_X^4}\right)^{1/2} 
\ee
The generation of leptonic asymmetry due to black hole evaporation producing heavy particles has some important features as mentioned in the introduction and corresponds to four different cases. The $T_{\mbox{evap}}$ as shown in Eq. (\ref{tevap}), could be less than or greater than $M_X$ having two possibilities. Also, the number of the $X$ particle produced by a single PBH has two possibilities as shown in Eqs.~(\ref{n<x}), and~(\ref{n>x}). Corresponding two different number densities of $X$ particle are obtained after multiplying these numbers by $n_{BH}$.

Combining all these, there are following four different cases: (\RN{1}) $T_{\mbox{evap}}\gtrsim M_{X}$ and  $T_{BH} > M_{X}$, (\RN{2})  $T_{\mbox{evap}} < M_{X}$ and  $T_{BH}>M_{X}$ (\RN{3})  $T_{\mbox{evap}} < M_{X}$ and  $T_{BH} \lesssim M_{X}$ (\RN{4})  $T_{\mbox{evap}}\gtrsim M_{X}$ and  $T_{BH} \lesssim M_{X}$. 
\begin{figure}[htb]
\centering
\includegraphics[scale=0.9]{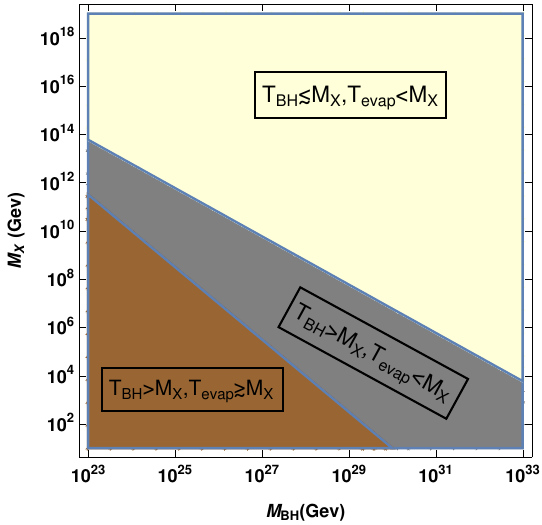}
\caption{The allowed region for $M_X$ versus $M_{BH}$ in cases \RN{1}, \RN{2}, and \RN{3} are shaded as brown, grey, and light yellow respectively.}
\label{allcase}
\end{figure}
Depending on the conditions on $T_{BH}$ and $T_{\mbox{evap}}$, we have shown the allowed region in $M_X- M_{BH}$ plane for cases I, \RN{2}, \RN{3}  in Fig.~\ref{allcase}. Mass $M_{BH}$ is restricted by conditions in Eq.~(\ref{T}) and Eq.~(\ref{up}). In Fig.~\ref{allcase}, we have varied the mass $M_X$ of the $X$ particle produced from PBH, in the range of $10$ GeV to about Planck scale. However, for case \RN{4},  there is no allowed region in $M-M_{BH}$ plane because for case \RN{4}, $T_{\text{evap}} \gtrsim M_{X}$ and $T_{BH} \lesssim M_{X}$. These imply  $M_{BH} \lesssim  6.5 \times 10^{18}\,\mathrm{GeV}$ which is not allowed by CMB mass bound as shown in  Eq.~(\ref{T}). Because of that, we will not consider this case for baryonic asymmetry. Fig.~\ref{allcase}  is not related to any specific model for lepton number violating interactions. However, in obtaining baryonic asymmetry, we will consider the supersymmetric model (MSSM) with heavy right-handed neutrinos for which also this Fig.~\ref{allcase} is valid.  
For case \RN{1}, from Fig.~\ref{allcase}, it is seen that for lower values of $M_X \sim 10 $ GeV , the allowed values of $M_{BH}$ are in the range of $10^{23} \mbox{GeV} \lesssim M_{BH} \lesssim 10^{30} \mbox{GeV} $. However, for higher values of $M_X$, the lower values of $M_{BH}$ are allowed.
 For case \RN{2}, from Fig.~\ref{allcase}, it is seen that for lower values of $M_X \sim 10$ GeV, only higher values of $M_{BH}$ are allowed in the range $10^{30} \mbox{GeV} \lesssim M_{BH} \lesssim 10^{33} \mbox{GeV} $. However, for higher values of $M_X \sim 10^{12}$ GeV, there could be lower values of $M_{BH}$ in the range of about $10^{23}-10^{25}$ GeV. For case \RN{3}, from Fig.~\ref{allcase}, one can see that for  $M_X > 10^{14}$ GeV, the entire range of $M_{BH}$ is allowed. For lower values of $M_X$ the higher values of $M_{BH}$ are allowed.

After PBH evaporation, for $T_{\mbox{evap}} \gtrsim M_X$ condition, the temperature of the universe is higher than the mass of the decaying particle $X$. So there is the possibility of the decaying particle $X$ to be in thermal equilibrium. In such cases with the expansion of the universe corresponding temperature of the universe will be lower and one is required to check in such cases whether the out-of-thermal equilibrium condition will be satisfied to get the leptonic asymmetry. 
 So for case \RN{1} only with condition $T_{\mbox{evap}} \gtrsim M_X$, (case   \RN{4} is already ruled out) the out of equilibrium condition 
\be\label{eql}
\Gamma_X \lesssim H (T = M_X),  
\ee
is to be satisfied, where $H$ is the Hubble's constant and $T$ is the universe's temperature.

\section{Baryonic asymmetry in MSSM with PBH and different constraints}
\label{sec:baryon}

\subsection{MSSM with heavy right-handed neutrinos and CP asymmetry }

In this section, we study the effect of heavy particle production from Hawking evaporation of PBH on leptogenesis. For that, we consider the Minimal Supersymmetric Standard Model (MSSM) with heavy right-handed neutrinos. The super-partner of such right-handed neutrinos has a lepton number violating decay mode. Because of that leptonic asymmetry and subsequently baryonic asymmetry could be produced. Apart from that, in the presence of such heavy right-handed neutrinos, a Type I see-saw mechanism for obtaining light neutrino mass could be implemented. 

For discussion on leptogenesis, we consider only one heavy RHN (which is the lightest among three RHNs) and its superpartner sneutrino. The mass term and the interaction terms related to RHN and sneutrino field, in the Lagrangian  can be written as 

\begin{eqnarray}\label{ln}
-{\cal L}_{\widetilde{N}} & = & M^{2}\widetilde{N}^{*}\tilde{N}
+\left(M^{*}Y_{\alpha}\widetilde{N}^{*}\widetilde{\ell}_{\alpha}H_{u}
+Y_{\alpha}\overline{\widetilde{H}_{u}^{c}}P_{L}\ell_{\alpha}\widetilde{N}+{\rm H.c.}\right),
\end{eqnarray}
where $P_{L,R}= \frac{1}{2}\left( 1\mp \gamma_{5}\right)$. 
Here $\tilde{N^c}$, $\tilde{\ell}_{\alpha}$ and $\tilde{H}_{u}$ represent chiral superfields of RHNs, the lepton doublet, and the up-type Higgs doublet, 
and $ \alpha$ is the lepton flavor indices. The soft supersymmetry breaking term is written as 

\begin{eqnarray}
-{\cal L}_{{\rm S}} & = & \widetilde{M}^{2}\widetilde{N}^{*}\tilde{N}
+\left(\frac{1}{2}BM\widetilde{N}\widetilde{N}
+A_\alpha \widetilde{N}\widetilde{\ell}_{\alpha}H_{u}+{\rm H.c.}\right).
\label{eq:soft}
\end{eqnarray}
The soft SUSY breaking parameters at electroweak/TeV energies have simplified forms at a usually high scale. Various restrictions on the soft parameters come from constraints from flavor physics, $CP$ violation, electroweak symmetry breaking, cosmology and collider physics \cite{Chung:2003fi, Nilles:1983ge,PhysRevD.27.2359}. It is useful to consider certain minimal framework for the pattern of soft parameters, for which $A_\alpha = a_0 Y_\alpha$. For the sake of simplicity, later on, we are considering that  $|Y_\alpha| \approx |Y| \approx Y $  and $A_\alpha=A$  and ignore the $\alpha$ index and restrict the maximal value of $A$ by considering $a_0 \lesssim \text{1 TeV}$.  In the minimal flavor violation scenario, $A$, $Y$, and $B$ could in principle be complex. If the complex phases in  $A$, $Y$, and $B$ are absorbed in the fields, then with minimal framework condition, it is found that there will remain a complex phase and without losing any generality, we may consider that phase in the tri-linear soft breaking parameter $A$.

The bilinear $B$ term in Eq. (\ref{eq:soft}), induces mixing in $\tilde{N}$ and $\tilde{N}^{*}$ to form mass eigenstates $\tilde{N}_{+}$ and $\tilde{N}_{-}$ with masses given by 
\be
M_{\pm}^{2} = M^{2}+\widetilde{M}^{2}\pm BM.
\label{eq:masses}
\ee
with the condition  $B < M + \widetilde M^2/M$ for real values of $M_\pm$. 
 
In the mass eigenstate basis, the Lagrangian is written as 
%Eq. (\ref{eq:mass_eigenstates}), 

\begin{eqnarray}
-{\cal L}_{\widetilde{N}}-{\cal L}_{{\rm S}} & = & 
M_{+}^{2}\widetilde{N}_{+}^{*}\widetilde{N}_{+}+M_{-}^{2}\widetilde{N}_{-}^{*}\widetilde{N}_{-}\nonumber \\
 &  & +\frac{1}{\sqrt{2}}\left\{ \widetilde{N}_{+}\left[ Y \overline{\widetilde{H}_{u}^{c}}P_{L}\ell
+\left(A + M Y\right)\widetilde{\ell}H_{u}\right]\right.\nonumber \\
 &  & \left.+i \widetilde{N}_{-}\left[ Y\overline{\widetilde{H}_{u}^{c}}P_{L}\ell
+\left(A- M Y\right)\widetilde{\ell}H_{u}\right]+{\rm H.c.}\right\} .
\label{eq:lag}
\end{eqnarray}

The amount of $CP$ asymmetry parameter $\eta_{\tilde{N}_{\pm}}$ for right-handed sneutrino  decay with final states $f=(\tilde{H_u} {l},{H_u} \tilde{l})$,
is written as
\begin{figure}
\includegraphics[scale=.8]{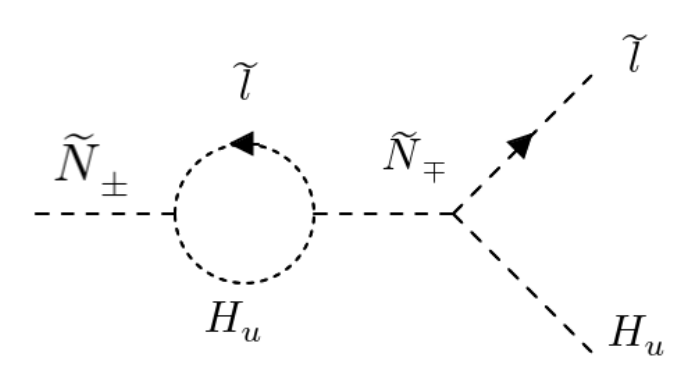}
\caption{One of the one loop self energy diagrams for $\tilde{N_{\pm}} \rightarrow \tilde{l} H_u$ with arrows indicating flow of lepton number} 
\label{oneloop}
\end{figure}
\begin{equation}
\label{genasym}
\eta_{\tilde{N_{\pm}}} \equiv  \frac{ \Gamma_{\tilde{N_{\pm}}}({\tilde{N_{\pm}}} \rightarrow {f}  )-\Gamma_{N_{\pm}}(\tilde{{N_{\pm}}}\rightarrow \bar{f} )}{\Gamma_{\tilde{N_{\pm}}}({\tilde{N_{\pm}}} \rightarrow {f}  )+\Gamma_{N_{\pm}}(\tilde{{N_{\pm}}}\rightarrow \bar{f} )},
\end{equation}
as $ \tilde{N_{\pm}} $ and its antiparticle are same.
Here total decay width of  $\widetilde{N}_\pm$ for tree level diagram is given by
\be
\Gamma_{\pm} \simeq \frac{M}{4 \pi}
 \left[Y^2 + \frac{|A|^2}{2M^2} 
\pm \frac{Y {\: \rm Re}(A)}{M}
%- \frac{Y_\alpha B\, {\rm Re}(A_\alpha) }{2M^2}
\right],
\label{eq:decay_width}
%\mp \frac{|A_\alpha|^2 B}{4 M^3} \right],
\ee
where for simplicity we have neglected the terms ${\tilde{M}^2}/{M^2}$ and ${B}/{M}$ assuming that $\tilde{M}<M$  and $B<<M$. We will set the condition that  $Y<1$ for the perturbative regime.

Non-zero $CP$ asymmetry $\eta_{\pm}$ is obtained from the interference of tree-level diagrams and self-energy one-loop diagrams (one loop vertex diagram gives a very small contribution to $CP$ asymmetry and has been neglected). 
  For  $ \tilde{N_{\pm}}  \rightarrow   \tilde{H_u} l $, there are two self-energy diagrams with $ H_u$  and $\tilde{l}$ in the internal lines. For $ \tilde{N_{\pm}}  \rightarrow   H_u \tilde{l} $, there are four self-energy diagrams,  in two of those- there are fermions- $ \tilde{H_u}$  and $ l $ in the internal line, and in other two of those, there are bosons- $ H_u$  and $ \tilde{l} $ in the internal lines. For illustration, one of the last two cases is shown in Fig.~\ref{oneloop} with the appropriate flow of the lepton number. In three of these self-energy diagrams, the lepton number arrows in the internal line are in the anticlockwise directions, for which non-zero $CP$ asymmetry can be obtained.  We have discussed earlier that  $A$ could be treated as complex coupling and one can see from Eq.~(\ref{eq:lag}), that the coupling of $\tilde{N_+}$ and $\tilde{N_-}$ with the slepton and Higgs are proportional to $A+MY$ and $A-MY$ respectively with different $CP$ violating phases in which $A$ may be considered as complex but $M Y$ is real. As the one loop diagram always contains both $\tilde{N_+}$ and $\tilde{N_-}$, the $CP$ violating phase will remain even if $A_\alpha$ is treated as universal $A \equiv A_\alpha$ parameter which has been considered in this work.  Considering the interference between tree level amplitude and one loop amplitude (with the non-zero imaginary part of the loop integral) in the numerator for both $f$ and $\bar{f}$ final states, one gets a non-zero numerator in Eq.~(\ref{genasym}), in the non-thermal case (with $T=0$). One may write the non-thermal $CP $ asymmetry \cite{Adhikari:2015ysa}  (after summing over all lepton flavors in the final states and also summing over all lepton flavors in the internal lines) as follows 
\bea
\label{eta}
\eta_{\tilde{N_{\pm}}}
& = & \frac{ 9 \: \rm Im (A) Y}{4\pi D_\pm M}
 \left( \frac{|A|^2 }{M}+\frac{\widetilde M^2 Y^2}{M} \pm {B Y^2}  \right)
\frac{4 B }{4 B^2 +\Gamma_\mp^2}
\eea 
where 
\be
\label{eq:D_T}
D_\pm \equiv \left[Y^2 +  \frac{|A|^2}{M^2} 
\pm \frac{2Y{\rm Re}(A)}{M}  \right] 
+ Y^2 \left(1 + \frac{\widetilde M^2}{M^2} \pm \frac{B}{M} \right).
\ee

After considering minimal flavor violation (to avoid flavor problem in Supersymmetry) with $A_\alpha = a_0 Y_\alpha$ alignment, and the Yukawa couplings due to one heavy right handed neutrino, the leptonic and baryonic asymmetry do not depend on complex phases present in $Y_D$.  However, if non-minimal flavor violation is considered and $A_\alpha$ is not aligned with $Y_\alpha$ then the complex phases in $Y_D$ matrix elements could lead to some extra contributions in the expression of $CP$ asymmetry. Without proper understanding of such phases, it is difficult to comment whether this could lead to either increase or decrease in $CP$ asymmetry depending on the phases associated with $Y_D$ matrix elements.

\subsection{Baryonic asymmetry with PBH and allowed parameter space}
 
For supersymmetric particles $\widetilde N_\pm$ produced due to the evaporation of a black hole, the leptonic asymmetry will be obtained from Eq.~(\ref{BHdom}), for $T_{BH}>M_X$ and from Eq.~(\ref{ydl}), for $ T_{BH} \lesssim M_X $
after replacing $\eta_{X}$ by $\eta_{\tilde{N_{\pm}}}$ as shown in  Eq.~(\ref{eta}). For $T_{BH} > {M_\pm}$ using Eqs.~(\ref{BHdom}), and ~(\ref{eta}), we can write 

\be\label{full}
Y_{\Delta L_\pm}\approx
\frac{  2.2 \times 10^{-2} f_X \:{g}^{1/4}   \: \rm Im (A)\: Y}{ \pi D_\pm M}
 \left( \frac{|A|^2 }{M}+\frac{\widetilde M^2 Y^2}{M} \pm {B Y^2}  \right)
\frac{4 B }{4 B^2 +\Gamma_\mp^2} 
\left( \frac{ M_P}{M_{BH}}\right)^{1/2}
\ee
where $Y_{\Delta L_\pm}$ correspond to the leptonic asymmetry due to $\tilde{N_+}$ decay and $\tilde{N_-}$ decay respectively. This expression of leptonic asymmetry is to be considered for case \RN{1} and case \RN{2}.  For $T_{BH} \lesssim  {M_\pm}$ using Eqs.~(\ref{ydl}), and ~(\ref{eta}), leptonic asymmetry is written as

\be\label{asymy}
Y_{\Delta L_\pm}\approx \frac{ 3.7 \times 10^{-5}   f_X \: g^{1/4} \: \rm Im (A)\: Y}{\pi D_\pm M}
 \left( \frac{|A|^2 }{M}+\frac{\widetilde M^2 Y^2}{M} \pm {B Y^2}  \right)
\frac{4 B }{4 B^2 +\Gamma_\mp^2} \left( \frac{ M_P^9}{M_{BH}^5 M_\pm^4}\right)^{1/2}.
\ee
This expression of leptonic asymmetry corresponds to case \RN{3} and case \RN{4}. But as discussed earlier, there is no allowed region in $M_X - M_{BH}$ plane for case \RN{4}. So this expression of asymmetry will be considered for case \RN{3} only.
Due to two different sets of conditions, Eqs.~(\ref{full}), and ~(\ref{asymy}), are the two expressions for leptonic asymmetry due to PBH evaporation. Here, asymmetry also depends on $M_{BH}$ (as sneutrinos are produced from PBH) apart from depending on $M, A, B$, and $Y$.

Leptonic asymmetry will result in the formation of baryonic asymmetry of the universe in the presence of shpeleron transition. Near the first order electroweak phase transition of the universe, $B+L$ violating spheleron transition becomes very small as the transition rate is Boltzmann suppressed and proportional to $e^{\frac{-E_{sph}}{T}}$ where $E_{sph}$ is the barrier height (related to up and down type Higgs vacuum expectation values). However, $B-L $ remains conserved.  Any change in $\Delta {L}$ will be related  to $\Delta {B}$ and the leptonic asymmetry $Y_{\Delta L}$ in either Eq.~(\ref{BHdom}), or Eq.~(\ref{ydl}), will be converted to baryonic asymmetry $Y_{\Delta B}$  \cite{saphos} :

\begin{align*}
Y_{\Delta B_{\pm}} = \frac{n_{B}-n_{\overline{B}}}{s} =  - \left( \frac{8 N_f + 4 N_H}{ 22 N_f + 13 N_H} \right) Y_{\Delta L_{\pm}}
\end{align*}
where $N_H$ is the number of Higgs doublets and $N_f$ is the number of lepton generations in the above equation. 
Considering $N_H =2$ and $N_f =3$ for MSSM, above relation is,
\be\label{sph}
Y_{\Delta B_{\pm}}=-\frac{8}{23}Y_{\Delta L_{\pm}}
\ee
In obtaining baryonic asymmetry from leptonic asymmetry in the presence of spheleron,  the weakly first-order or second-order phase transition is required. In MSSM this transition could happen at the temperature of the universe $T \gtrsim 200 $ \rm  GeV \cite{hambay2000,Rubakov1996,riotto,moore1999}. Using Eq. (\ref{tevap}) the PBH mass could be less than about $1.79 \times 10^{29}$ GeV corresponding to such $T \approx T_{evap}$. In obtaining the appropriate baryonic asymmetry, we have used this constraint due to sphaleron transition on $M_{BH}$  to show the allowed regions for different parameters in Fig.~\ref{short} later. However, without the constraint due to sphaleron transition on $M_{BH}$, we have also shown the enlarged parameter space in Fig.~\ref{1case2} for which leptonic asymmetry will be generated, however, conversion of that asymmetry to baryonic asymmetry, would not be possible.  

The Yukawa matrix $Y_D$ corresponding to $Y_\alpha$ in Eq.~(\ref{ln})  is given in terms of the relevant low-energy observables under the Casas-Ibarra (CI) parameterization \cite{Casas:2001sr} as follows \cite{Morisi:2024yxi}:
\bea\label{yll}
Y_D  = v_u^{-1} U_{PMNS} \sqrt{m_\nu^{diag}}
R \sqrt{M_N^{diag}}
\eea
where $R$ is an arbitrary orthogonal matrix which could be complex provided that $R^T R = 1$. Here, $v_u$ is the vev of $H_u$ field in Eq. (\ref{ln}), $U_{PMNS}$ is the light neutrino mixing matrix, $m_\nu^{diag}$ is the diagonal light neutrino mass matrix and $M_N^{diag}$ is the diagonal heavy right-handed neutrino mass matrix. The Yukawa couplings could be of order one even for the heavy right-handed neutrino masses $M$ in the GeV range, for large imaginary parts in the matrix elements of $R$ as discussed in ref. \cite{Morisi:2024yxi}.
 $Y_D$  is, in general, a complex matrix.  In the context of soft leptogenesis we like to make a few comments in this regard. The leptonic asymmetry  is generated through the decay of sneutrinos.  The soft breaking terms involving the singlet sneutrinos remove the mass degeneracy of two real sneutrinos state of single heavy neutrino generation. The mixing between two sneutrinos states generate $CP$ asymmetry in the decay. We have studied the soft leptogenesis due to one heavy right handed neutrino (which is the lightest one among three heavy right handed neutrinos)  and its super-partner. With this consideration, three Yukawa couplings $Y_\alpha$ which are in general complex, will play role in the leptonic asymmetry. However, based on Eqs.~(\ref{ln}) and (\ref{eq:soft}), the leptonic asymmetry  will depend on one non-flavor phase as we consider the alignment of soft breaking trilinear $A_\alpha $ with $Y_\alpha$ as $A_\alpha = a_0 Y_\alpha$, where  $a_0$ is around TeV scale. For soft leptogenesis, we have considered the lightest right-handed neutrino and its super partner. However, in Type I seesaw mechanism, with the usual three right-handed neutrinos, there are more Yukawa couplings present in the tree level Dirac mass matrix $M_D = Y_D v$ and at the tree level $M_D$ does not depend on trilinear soft breaking term. So while using Eq.~(\ref{yll}) in connection with light neutrino masses and mixing, $Y_D$ is the usual complex matrix. In considering soft leptogenesis, where soft terms are also playing role, complex phases in some of the Yukawa couplings, could be absorbed after assuming that $A$ parameter is aligned to $Y$ parameter.

In obtaining observed baryonic asymmetry in Eq.~(\ref{first}), using Eq.~(\ref{asymy}) or (\ref{sph}), the large Yukawa couplings are required. In our analysis, the  Yukawa couplings as small as $0.01$ have been considered but appropriate asymmetry is obtained for $Y$ about   $0.06$  and above and discussed in the context of the allowed region of different parameters in Fig~\ref{short}. In MSSM,
\bea\label{cot}
v_u^2+v_d^2= v_u^2 (1+ \cot^2 \beta) \simeq (174 \; \rm GeV)^2
\eea
where $v_u,v_d$ =$\langle H_{u,d} \rangle $ are up-type and down-type Higgs vacuum expectation values. $M$ which is the lightest among the heavy RHN masses in matrix $M_N^{diag}$ in Eq. (\ref{yll}), which has been considered as the approximate quasi-degenerate mass scale of three heavy right-handed neutrinos and $m_{\nu_i}$  are the three active light neutrino masses in $m_\nu^{diag}$ in Eq.~(\ref{cot}), the sum of which is   $\lesssim 0.118 \: \rm eV $~\cite{RoyChoudhury:2018gay} due to cosmological upper bound on light nuetrino masses, $ \rm tan \beta$$ \equiv v_u/v_d $. Due to the lighter neutral Higgs mass ( $125.35$ GeV) found at LHC, $\tan \beta > 4$ is expected \cite{Djouadi}.  

In the context of leptogenesis in the supersymmetric model, $M_X$ stands for sneutrino mass $M_\pm$ which is related to right-handed neutrino mass $M$ as shown in Eq.~(\ref{eq:masses}). In Fig.~\ref{allcase} we have shown a variation of $M$ with other parameters.
Masses of sneutrinos $\widetilde N_+$ and $\widetilde N_-$ do not differ much and $M_{\pm} \sim M$ for higher values of $M$ in comparison to $B$.  In the figures below, we have considered the decay of $\widetilde{N}_{-}$ only, the mass of which is lighter than that of $\widetilde{N}_{+}$ as well as right-handed neutrino mass $M$. For simplicity, the mass parameter $\widetilde{M}$ has been assumed to be zero in evaluating baryonic asymmetry in the latter part. 

For case \RN{1} only, the out-of-equilibrium condition is to be satisfied. Using Eq.~(\ref{eq:decay_width}), in Eq.~(\ref{eql}), for the sneutrino decay,
this condition is obtained as 
\be
\frac{\left|\rm A\right| ^2}{2 \rm M^2}\pm \frac{\rm Y \rm\: Re[A]}{\rm M}+\rm Y^2 \lesssim \frac{4 \pi}{M} H_{T \sim M} \lesssim 3.04 \times  10^{-10}  \left( \frac{M}{10^7 \, GeV} \right)
\label{eq:out-of-equilibrium}
\ee
where the Hubble expansion rate $H$ is given by $H= 1.66\sqrt{g}\; T^2/M_{P} $ with Planck mass $M_{P} = 1.22 \times 10^{19}$ GeV and $g\approx 316$ \cite{Keith2016}. Considering $ \widetilde{M} $ and $ B $ much smaller than $ M $ in Eq. (\ref{eq:masses}), we have considered $ M_{\pm} $  of the order of $M$ and hence $T \sim M $ has been considered in Eq.~(\ref{eq:out-of-equilibrium}).
For case \RN{1}, this condition will constrain parameters $A$ and $Y$ depending on $M$ values.

\begin{figure}
\centering
\begin{tabular}{cc}
\includegraphics[scale=0.7]{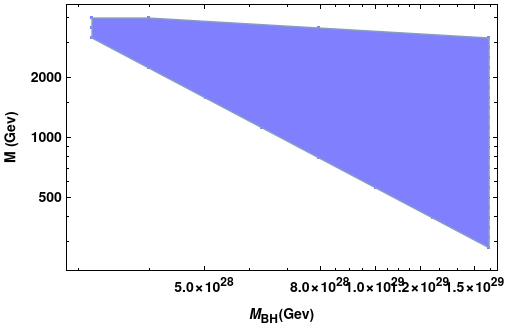}&
\includegraphics[scale=0.7]{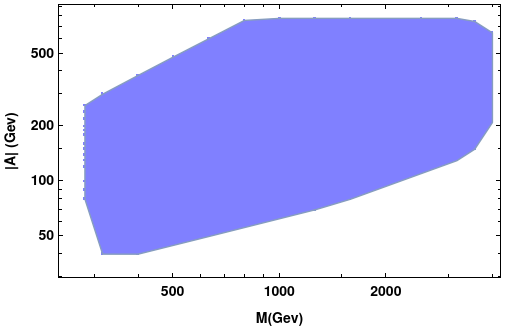}\\
\bf{(a)} & \bf{(b)} \\
\includegraphics[scale=0.7]{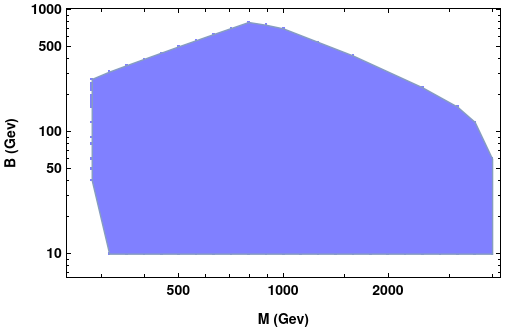}&
\includegraphics[scale=0.7]{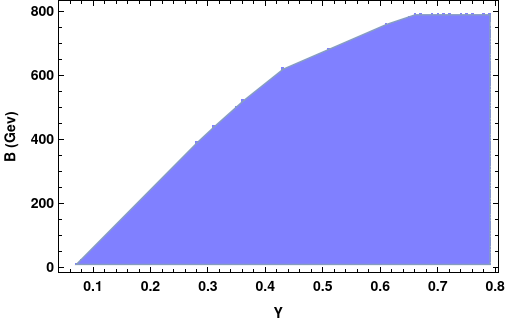}\\
\bf{(c)} & \bf{(d)} \\
\includegraphics[scale=0.7]{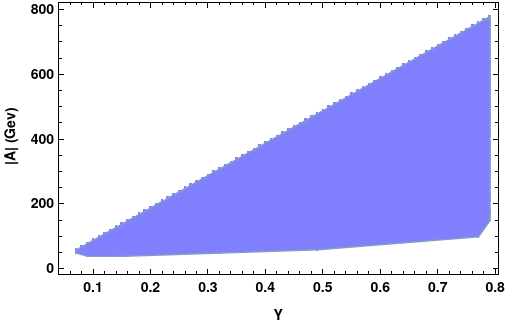}&
\includegraphics[scale=0.7]{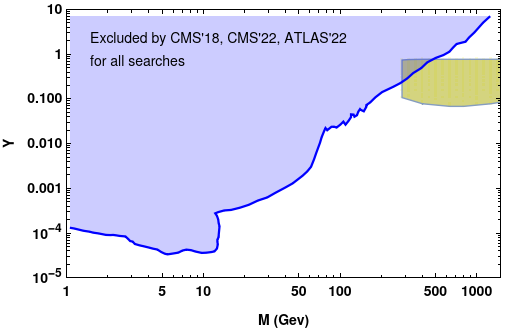}\\
\bf{(e)} & \bf{(f)} \\
\end{tabular}
\caption{The allowed regions in shaded blue for $M$ versus $M_{BH}$, $|A|$ versus $M$, $Y$ versus $B$, and $Y$ versus $|A|$ for spheleron bound, are shown respectively in (a),(b),(c), (d) and (e) for case \RN{2}. In panel (f), the allowed region of $Y$ versus $M$ is shaded in light yellow outside the blue-shaded disallowed region constrained by combined bounds of CMS '18, CMS '22, and ATLAS '22.}
\label{short}
\end{figure}

\begin{figure}
\centering
\begin{tabular}{cc}
\includegraphics[scale=0.7]{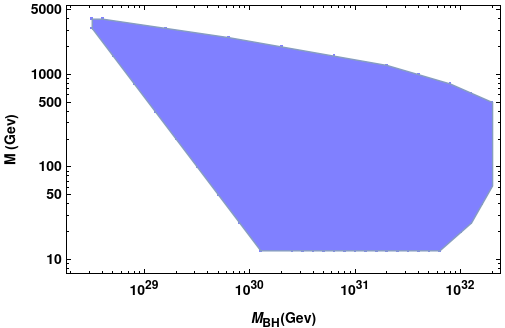}&
\includegraphics[scale=0.7]{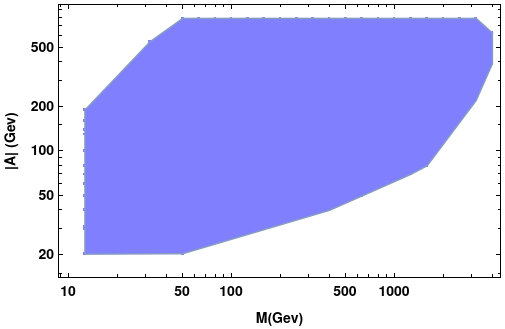}\\
\bf{(a)} & \bf{(b)} \\
\includegraphics[scale=0.7]{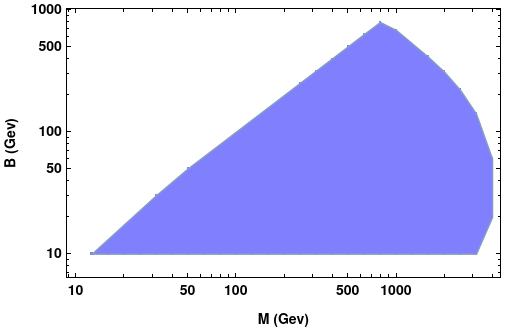}&
\includegraphics[scale=0.7]{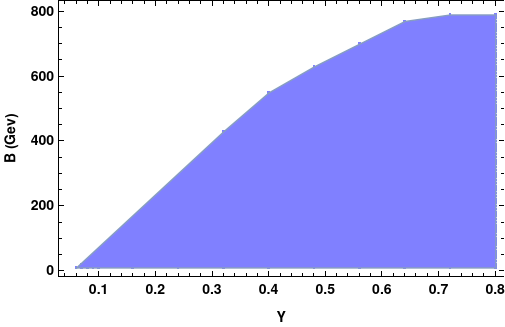}\\
\bf{(c)} & \bf{(d)} \\
\includegraphics[scale=0.7]{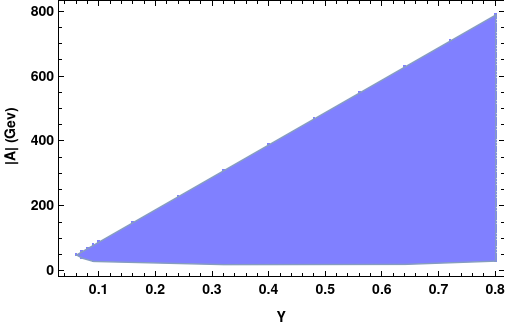}&
\includegraphics[scale=0.7]{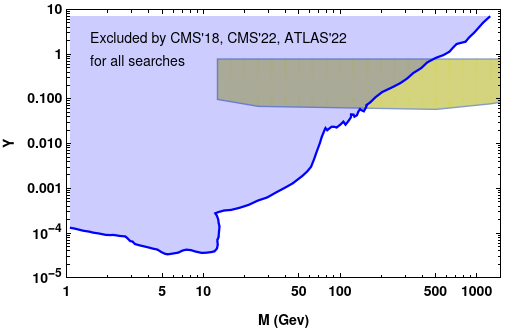}\\
\bf{(e)} & \bf{(f)} \\
\end{tabular}
\caption{The allowed regions in shaded blue for $M$ versus $M_{BH}$, $|A|$ versus $M$, $Y$ versus $B$, and $Y$ versus $|A|$ are shown respectively in (a),(b),(c), (d) and (e) for case \RN{2}. In panel (f), the allowed region of $Y$ versus $M$ is shaded in light yellow outside the blue-shaded disallowed region constrained by combined bounds of CMS '18, CMS '22, and ATLAS '22.}
\label{1case2}
\end{figure}

Following the earlier discussion  below Eq. (\ref{eq:soft}), 
to find the suitable values of soft supersymmetry breaking parameters that give appropriate leptonic and baryonic asymmetry, we have varied the parameters $M$ and $M_{BH}$ related to the masses of sneutrinos and PBH masses. From Eqs.~(\ref{T}), and ~(\ref{up}), the valid range of   $M_{BH}$  is considered to be from about ($1.49 \times 10^{23}$ to $ - 6.1 \times 10^{32} $) \mbox{GeV}. However, for getting baryonic asymmetry, in Fig.~\ref{short}, we have used the sphaleron transition constraint as $M_{BH} \lesssim 1.79 \times 10^{29}$ as discussed earlier after Eq.~(\ref{sph}). However, in Fig.~\ref{1case2} as we have considered only the generation of leptonic asymmetry only, the upper bound on $M_{BH}$ is $6.1 \times 10^{32} $. 
There are no specific lower or upper bounds on the masses of heavy right-handed neutrinos and it spans any value between the fraction of eV up to the GUT scale \cite{Abdullahi:2022jlv}. Our analysis has varied it from about $ 10 $ GeV to $10^{10}$ GeV.  Furthermore, as stated after Eq.~(\ref{eq:soft}), we consider the soft trilinear parameter $|A|$ to approximately follow $A \lesssim Y \times \mbox{TeV scale}$. The other soft bilinear parameter $B$ is also considered around the electroweak scale to the TeV scale for the naturalness in MSSM. 
We show that if sneutrinos are produced from PBH then it is possible that the observed baryonic asymmetry could be produced for the natural choice of the soft SUSY parameters. With full parameters scan of $|A|$, $B$, and $Y$,  we show the possible allowed regions of the soft SUSY parameter space with the lightest right-handed neutrino mass $M$  and also the allowed region of PBH mass ($M_{BH}$) with the corresponding $M$ value in Figs.  \ref{short} and \ref{1case2} with and without sphaleron transition constraint on $M_{BH}$ respectively.

For  
 case (\RN{1}) with conditions $T_{\mbox{evap}}\gtrsim M_{X}$ and $T_{BH} > M_{X}$ and case (\RN{3})  with conditions $T_{\mbox{evap}} < M_{X}$ and  $T_{BH} \lesssim M_{X}$, we have checked that for soft parameters around the electroweak scale to TeV scale, it is not possible to obtain appropriate baryonic asymmetry . For case 
 (\RN{1}), as discussed earlier, the out-of-equilibrium condition in Eq. (\ref{eq:out-of-equilibrium}) is to be satisfied. For that Yukawa coupling $Y$ is required to be very small in the range of about $10^{-7}$ or smaller depending on $M$ value about $10^3$ GeV or lesser. For case 
 (\RN{1}), the expression of baryonic asymmetry to be considered is given in Eq. (\ref{full}) and as this asymmetry is overall proportional to $Y$, apart from other suppression factors, it is not possible to get asymmetry from case 
 (\RN{1}). For case (\RN{3}), although the out-of-equilibrium condition is not required and $Y$ value may not be suppressed like case (\RN{1}), however, the second expression of baryonic asymmetry given in Eq. (\ref{asymy}) which is to be considered, is highly suppressed by the higher powers of allowed $M$ and $M_{BH}$ (from Fig.~\ref{allcase} for case (\RN{3})) in the denominators. For this reason, it is not possible to obtain the required baryonic asymmetry for case (\RN{3}).   Case (\RN{4}) is not possible as discussed earlier. However, for case (\RN{2}) with conditions $T_{\mbox{evap}} < M_{X}$ and  $T_{BH}>M_{X}$, we find that for natural values of soft parameters, it is possible to obtain asymmetry from the expression given in Eq. (\ref{full}). In this case, the out-of-equilibrium condition is not required to be satisfied and the higher values of $Y$ could be considered. Furthermore, the leptonic asymmetry expression to be considered is given in Eq.~(\ref{full}) which is not so much suppressed by $M$ and $M_{BH}$ unlike case (\RN{3}) in Eq.~(\ref{asymy}).  

In Fig. \ref{short},  we discuss the case (\RN{2}) in detail for which appropriate baryonic asymmetry is possible with soft SUSY breaking parameters near their natural electroweak values. In plotting Fig. \ref{short}, we have considered the relationship of $|A|$ and $Y$ as  $|A| \lesssim 10^3 Y$  GeV, and the value of $B$ is nearer the electroweak scale for naturalness. Also, we have considered  $M>B$ to avoid an un-physical tachyonic solution of $M$. In Fig \ref{short} (a),  the dark green shaded region is shown to be the allowed region for obtaining appropriate baryonic asymmetry. In the allowed region
$M_{BH}$ is found to be  from about $3.16 \times 10^{28}$ GeV to $ 1.79 \times 10^{29}$ GeV and $M$ is found to be from about $300 \mbox{GeV}$ to about $4$ TeV. For lower values of $M$ the higher values of $M_{BH}$ are only possible.  
 In Fig. \ref{short} (b) the allowed blue-shaded region of $|A|$ versus $M$ values are shown. For low values of $|A|$ around $40$ to $50$ GeV, $M$ values are also lower from about $300$ to about $500$  GeV. However, for $M$ above $500$ GeV, $|A|$ value could be from about $40$ GeV to about $800$ GeV. In Fig.\ref{short} (c), the allowed blue-shaded region of $B$ versus $M$ is shown. For high values of $B$ around $800$ GeV, the $M$ value is around $900$ GeV. For lower values of $B$, the $M$ value could be in the range of about 300 GeV to $4$ TeV.  In Fig.\ref{short} (d) and (e) we have shown the blue-shaded allowed region of $B$ versus $Y$ and $|A|$ versus $Y$ respectively. The minimum value of $Y$ is around $0.06$. However, we have restricted the higher values $Y$ to $0.8$. From Fig.\ref{short} (d) and (e), it is found that the higher values of $B$ and $|A|$ are possible for higher values of $Y$.

In Fig.\ref{short} (f), we have shown the allowed region of $Y$ versus $M$ in the light yellow-shaded region for obtaining baryonic asymmetry. The light blue-shaded region is the excluded region obtained from different CMS, ATLAS experimental constraint \cite{Abada:2022wvh} on collider dilepton, trilepton, and long-lived searches for heavy neutral lepton which is heavy right-handed neutrino in our case.  Considering the relation of mixing with the Yukawa couplings as $|U_{\alpha i}|^2 \sim |Y_{ij} v_u /M_j|^2$ \cite{Drewes:2019mhg} for diagonal $M_R$ in the seesaw mass matrix and $Y_{ij} \equiv Y$ and $M_j \equiv M$, 
we have interpreted the most stringent experimental bounds on mixing versus $M$
in terms of  the Yukawa coupling $Y$ versus $M$ in Fig.\ref{short} (f). After using this experimental constraint, it is found that although the lower bound on $M$ is still around $300$ GeV, however, for that mass, the Yukawa coupling is required to be lesser than about $0.4$. However, for those values of $Y$ baryonic asymmetry is still possible. However, for $M$ values above 500 GeV, there is no further constraint coming from collider searches.
So in obtaining baryonic asymmetry through sneutrino decay in MSSM, the lower bound of right-handed neutrino mass may be considered as about $300$ GeV.

Next Fig. \ref{1case2} is shown like Fig. \ref{short}, however, without imposing a constraint on $M_{BH}$ due to the requirement of suitable sphaleron transition. But the upper bound on $M_{BH}$ due to BBN, has been considered. Because of this, in comparison to Fig. \ref{short}, the overall parameter space has increased in all the cases of Fig. \ref{1case2} and the $M$ value could be further lower. Tables \ref{tab:parameter1} and \ref{tab:parameter2} show the allowed ranges of $M_{BH}$, $M$,$A$,$B$, and $Y$ for successful leptogenesis with sphaleron bound and without sphaleron bound respectively.
\begin{table}
\centering
\begin{tabular}{|c|c|c|c|c|}
\hline
$M_{BH} $ (GeV) & M (GeV) & $|A|$ (GeV)& B (GeV) & Y \\ \hline
$3.2 \times 10^{28}- 1.5 \times 10^{29}$ & $300-3000$ & $50-800$ & $10-1000$ & 0.06-0.8 \\ \hline
\end{tabular}
\caption{Allowed range for successful leptogenesis  of  $M_{BH}$, M, $|A|$, B, Y.} with sphaleron bound 
\label{tab:parameter1}
\end{table}
\begin{table}
\centering
\begin{tabular}{|c|c|c|c|c|}
\hline
$M_{BH} $ (GeV) & M (GeV) & $|A|$ (GeV)& B (GeV) & Y \\ \hline
$3.2 \times 10^{28}- 1.1 \times 10^{32}$ & $200-1300$ & $20-800$ & $10-1000$ & 0.06-0.8 \\ \hline
\end{tabular}
\caption{Allowed range for successful leptogenesis  of  $M_{BH}$, M, $|A|$, B, Y.} without sphaleron bound 
\label{tab:parameter2}
\end{table}
\subsection{Constraints from Charged Lepton Flavor Violating (CLFV) process}

The soft supersymmetry breaking parameters can contribute to CLFV interactions, dipole moments of leptons, etc. The experimental upper bound on the branching ratio from MEG \RN{2} collaboration \cite{MEGII:2023ltw}
 \be\label{br}
 BR(\mu \rightarrow e \gamma) < 3.1 \times 10^{-13}
 \; .\ee
 This constrains the trilinear soft supersymmetry breaking parameter $|A|$ as discussed below. The above process is induced by one-loop diagrams through the exchange of gauginos (neutralinos, charginos) and sleptons (charged sleptons, sneutrinos). The main contribution comes from charginos and sneutrinos in the internal lines of the one-loop diagram. The off-diagonal elements of doublet slepton mass square $m_{\tilde l}^2$ can induce the CFLV process as considered in Eq.~(\ref{gut}), below.
 It is normally assumed as mSUGRA 
 \begin{figure}
\centering
\begin{tabular}{cc}
\includegraphics[scale=.7]{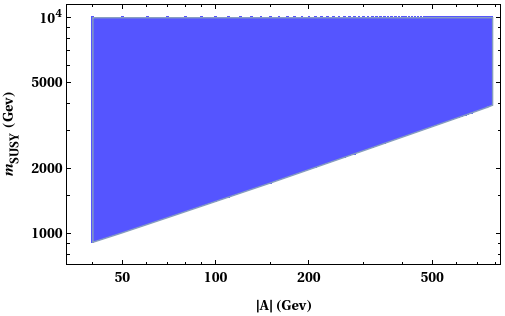}&
\includegraphics[scale=.7]{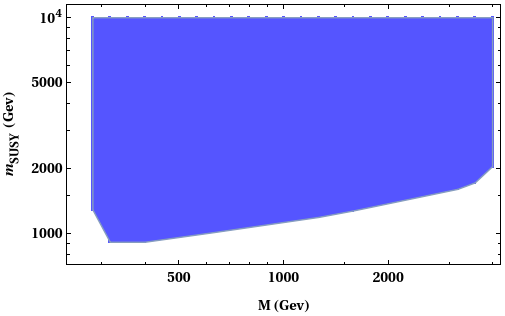}\\
\bf{(a)}&\bf{(b)}\\
\end{tabular}
\caption{Allowed region for $m_{SUSY}$ versus $|A|$ for spheleron bound, is shown in the shaded blue obtained from experimental constraint on branching ratio of $\mu \rightarrow e \gamma$.}
\label{clfvshort}
\end{figure}

 \begin{figure}
\centering
\begin{tabular}{cc}
\includegraphics[scale=.7]{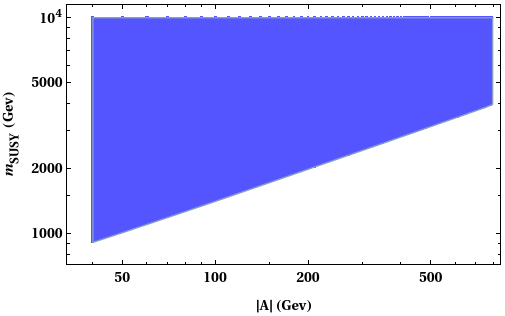}&
\includegraphics[scale=.7]{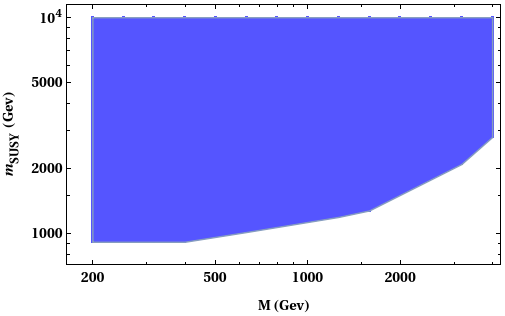}\\
\bf{(a)}&\bf{(b)}\\
\end{tabular}
\caption{Allowed region for $m_{SUSY}$ versus $|A|$, is shown in the shaded blue obtained from experimental constraint on branching ratio of $\mu \rightarrow e \gamma$.}
\label{clfv}
\end{figure}
boundary conditions that off-diagonal $m_{\tilde l}^2$ is zero at the GUT scale. It evolves from GUT scale $M_{\rm GUT}$ to the RHN mass scale $M$ through Renormalization Group Equations and   in the leading log approximation,  may be written (considering the contribution due to soft breaking $|A|$ term) as \cite{Hisano:1995cp}
 \be\label{gut}
 \left(m_{\tilde l}^2\right)_{\mu e} \approx -\frac{1}{8 \pi^2} A_\mu^* A_e \log \left( \frac{M_{\rm GUT}}{M}\right) \approx  -\frac{1}{8 \pi^2} {|A|}^2 \log \left( \frac{M_{\rm GUT}}{M}\right) 
 \; .\ee
The right-hand side expression follows as we have assumed $A_\alpha \equiv A$. Due to the lack of specific understanding of the hierarchical pattern of other heavier right-handed neutrinos, their similar contributions have been ignored here. The branching  ratio for $\mu \rightarrow e \gamma$ in terms of above $\left(m_{\tilde l}^2\right)_{\mu e}$ can be written as \cite{Hirsch:2012ti}
 \be\label{susy}
 BR ( \mu \rightarrow e \gamma ) \approx \frac{\alpha^3}{G_F^2} \frac{{|\left(m_{\tilde l}^2\right)_{\mu e}|}^2}{{m_{SUSY}}^8} \tan^2\beta
 \ee
where the fine structure constant at $m_W$ scale is  $\alpha \sim 1/128$ and the Fermi four point coupling $G_F \sim 1.166 \times 10^{-5}$ GeV$^{-2}$ and $m_{SUSY}$ is the typical mass scale of the supersymmetric particles.

Using the allowed region of $|A|$ versus $M$ as discussed in the context of Fig.~\ref{short} (b)  as well as taking into account that $M \gtrsim 300$ GeV is allowed as follows from Fig.~\ref{short} (f)   
and considering $\tan \beta = 10$ \cite{Djouadi} in Eq.~(\ref{susy}), and considering $M_{GUT} \sim 10^{16}$ GeV in Eq.~(\ref{gut}), one can find the allowed region of $m_{SUSY}$ versus $|A|$ and $m_{SUSY}$ versus $M$, as shown in
Fig.~\ref{clfvshort} (a) and (b) respectively subject to the experimental constraint on the branching ratio as shown in Eq.~(\ref{br}). From Fig.~\ref{clfvshort} (a) and (b), it is found that lower bounds of $m_{SUSY}$ are possible depending on $|A|$ and $M$ values.
For lower values of $m_{SUSY}$ around $1$ to $1.5$ TeV, the trilinear $|A|$ parameter is expected to be around $40$ to $150$ GeV and $M$ value around $300$ GeV to $2$ TeV. Without using sphaleron constraint on $M_{BH}$ and using the allowed region of $|A|$ versus $M$ as discussed in the context of Fig.~\ref{1case2} (b)  as well as taking into account that $M \gtrsim 160$ GeV is allowed as follows from Fig.~\ref{1case2} (f), Fig.~\ref{clfv} has been shown.

MEG II experiment could further improve the bound on the branching ratio of $\mu \rightarrow e \gamma$ to about $ 6 \times 10^{-14 }$ in the near future \cite{MEGII} and better statistics
 is foreseen by 2026.
 In that case, the lower bounds of $m_{SUSY}$ shown in Figs.~\ref{clfvshort} and ~\ref{clfv}, may be further higher depending on $|A|$ values.

In the analysis of Eqs. (\ref{gut})-(\ref{susy}), the universality of $A$ parameter at the GUT scale, has been considered. However, in general Supergravity models and in Type I string models, it is possible to have non-universality in the scalar masses, $A$-terms, and
gaugino masses \cite{PhysRevResearch.1.033022,Ellis:2016nke,Carvalho:2001ex}.
In such cases, the trilinear $A$ parameter could be written as $A_{ij} = a_{ij} Y_{ij}$ instead of universality condition $A_{ij} = a_{0} Y_{ij}$. Also $A_{ij}$ may not be related to $Y_{ij}$ \cite{Papadopoulou:2002}. In such cases, after going to the basis of diagonal Yukawa matrices through super-field rotations, there could be large off-diagonal terms in the trilinear couplings and large off-diagonal terms in the slepton mass matrices. 
This could lead to large flavor violating effects (which is the SUSY flavor problem). There could be some changes in the allowed region of $A$ parameter versus $m_{SUSY}$ as shown in Figs.\ref{clfvshort}(a) and \ref{clfv}(a), depending on the consideration of non-universality in $A$ parameters.

\section{Constraints on the mass of gravitino produced from PBH}
\label{natural}
In SUSYGUT theories \cite{Georgi:1974,GellMann:1978,Fritzsch:1975,Mohapatra:1999,Raby:2004} Gravity mediated supersymmetry breaking is possible in the hidden sector through Super-Higgs mechanism. This could give mass to gravitino around the weak scale.
Gravitino, a spin-3/2 super-partner of the graviton,  has interaction strength, with the observable sector - the standard model particles and their superpartners, are inversely proportional to the Planck mass. Unstable gravitinos could be abundant during nucleosynthesis and could destroy the good agreement of BBN theory with observations. If the gravitinos are the lightest supersymmetric particles and are stable, then also a constraint on their mass could come from the observed density parameter for dark matter. Such cosmological constraints for unstable and stable gravitinos were discussed earlier when they could be produced thermally. This subsequently leads to the constraint on reheating temperature. This restricts the mass of heavy particles (which are also thermally produced)  whose lepton or baryon number-violating decays are essential for the generation of baryonic asymmetry. In that way, cosmological constraints on gravitino constrains the scenario of baryogenesis. Here, we discuss such cosmological constraints on gravitino masses when gravitino is produced by the evaporation of PBH. With such production, we have shown in this section what could be the allowed region of gravitino mass versus PBH mass. In our discussion on baryogenesis through sneutrino decays, we have shown in Fig. \ref{tbh12}, the allowed region of PBH mass $M_{BH}$ and the right-handed neutrino mass $M$ (which is related to mass $M_\pm$ of decaying sneutrinos). Thus constraint on gravitino mass could be related to baryogenesis through $M_{BH}$ in our work.

Following Eq. (\ref{n<x}) and Eq. (\ref{n>x}), the number density $n_{3/2}$ of gravitinos produced by PBH evaporation (which is almost instantaneous as discussed in Section II) is given by

\be\label{gravpro}
n_{3/2}=
\begin{cases}
\displaystyle
\frac{n_{BH} \, 4 \pi \; {f_{3/2}} \; {M_{BH}}^2} { 3 \, M_{Pl}^2} & \text{if } m_{3/2} < T_{BH}, \\ 
 \\
\displaystyle
\frac{n_{BH} f_{3/2}  \;{ M_{Pl}}^2}{48 \pi  \; {m_{3/2}}^2} \; & \text{if } m_{3/2} \gtrsim T_{BH}
\end{cases}
\ee
at temperature $T = T_{evap}$ where $f_{3/2}\sim g_{3/2}/g$ and $g_{3/2}$ is the number of degrees of freedom for gravitino and $n_{BH} = \rho_{BH}/M_{BH}$ is obtained from Eq. (\ref{TRH1}). 

\subsection{Unstable gravitino}

 If gravitino is unstable then it could decay into some supersymmetric particles lighter than this.  
 The decay width of gravitino for all MSSM particles in the final states (after neglecting the mass of final states particles with respect to gravitino mass) is \cite{Nakamura:2006}
\[
\Gamma_{3/2} = \frac{193}{384\pi} \frac{m_{3/2}^{3}}{M_{Pl}^2} 
\]
One may write the thermally averaged decay width  as  
\[
\Gamma_{3/2}(z) = \Gamma_{3/2} {K_1(z) \over K_2(z)}
\]
Here $K_1(z)$ and $K_2(z)$ are modified Bessel functions of the first and second kinds, respectively.
\begin{figure}[htb]
\centering
\includegraphics[scale=.9]{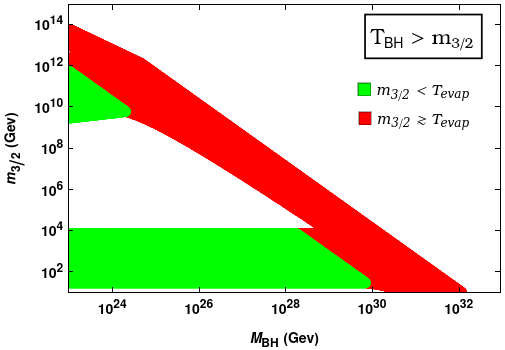}
\caption{The allowed region of $m_{3/2}$ versus $M_{BH}$ shaded in both green and red for unstable gravitino. The green and red correspond to $ m_{3/2} < T_{evap}$ and $m_{3/2} \gtrsim T_{evap} $ respectively. }
\label{tbh12}
\end{figure}

Using the Boltzmann equation,  the abundance of the unstable gravitinos at the BBN temperature of the universe could be obtained. Let us define  $z$ as the ratio of gravitino mass $m_{3/2}$ to the temperature $T$ of the universe (which is varying from $T_{evap}$ to $T_{BBN}$) and   $Y_{3/2}$ as ratio of the number density $n_{3/2}$ to entropy density $s(z)$ as given below : 
\[
z={m_{3/2} \over T};  \;\;\;\;\;\;\;  s(z)=\frac{2 \pi ^2  m_{3/2}^3 }{45 z^3};  \;\;\;\;\;     Y_{3/2}(z)={n_{3/2} \over s(z)};
\]
The evolution of $Y_{3/2}$ with $z$ is given by the Boltzmann equation as
\begin{equation}\label{Boltz}
\frac{dY_{3/2}(z)}{dz} = -{ Y_{3/2}(z) \over s(z) \; H(z) \; z}\left({Y_{3/2}(z) \over Y_{3/2}^{eq(z)}}-1 \right) \Gamma_{3/2}(z)
\end{equation}
where $Y_{3/2}^{eq}(z)$ is the thermal  equilibrium value of $Y_{3/2}(z)$ corresponding to fermionic number density.

In solving the Boltzmann  Eq.~(\ref{Boltz}), the following initial condition on $Y_{3/2}$ at $z= z_{evap}$ has been used  by using Eq.~(\ref{gravpro}) for two different cases, as 
\be\label{yevap}
Y_{3/2}(z_{evap})=
\begin{cases}
\displaystyle
{{n_{BH} \, 4 \pi \; {f_{3/2}} \; {M_{BH}^2} \over {3 \, M_{Pl}^2} \; s(z_{evap})}} & \text{for } m_{3/2} < T_{BH} \\
\\
\displaystyle
{{n_{BH} f_{3/2}  \;{ M_{Pl}}^2}\over{48 \pi  \; {m_{3/2}}^2} \; s(z_{evap})} & \text{for } m_{3/2} \gtrsim T_{BH}. 
\end{cases}
\ee
The photon dissociation of $^4He$ produces $D$ and $^3He$. As the mass abundance of $^4He$ is very large in comparison to $D$ and $^3He$, so a very small fraction of $^4He$ could be photo-disintegrated. This puts a stringent constraint on the upper bound of the primordial gravitino abundance at the BBN epoch and is given by \cite{Ellis:1984er}
\begin{equation}\label{gbound}
Y_{3/2}(z_{BBN})< \frac{3.2 \times 10^{-12}}{\pi^4} \left( \frac{m_{3/2}}{100 \, \text{GeV}} \right)^{-1}.
\end{equation}

From Eq.~(\ref{Boltz}),  $Y_{3/2}(z_{BBN})$  is obtained for different values of $m_{3/2}$ and $M_{BH}$ and using the constraint in Eq.~(\ref{gbound}), the allowed region of $m_{3/2} $ and $M_{BH}$ has been plotted in Fig.~\ref{tbh12}. The allowed green region corresponds to $m_{3/2} < T_{evap}  $ and the allowed red region corresponds to $ m_{3/2} \gtrsim T_{evap}$.  Based on some SUSY parameters, the experimental lower bound on the mass of the lightest supersymmetric particle (LSP) is $\gtrsim 30 \mbox{GeV}$ for neutralino \cite{Calibbi:2014coa},  For LSP other than neutralino, the lower bound on their masses is relatively higher in the range of about $650 - 2000$ GeV for  Higgsino, stau, stop, higgsino, etc., as LSP \cite{Heisig:2013sva,kpatcha2022searching,barman2020status}. So depending on the type of LSP, in Fig.~\ref{tbh12}, the lowest possible values of unstable $m_{3/2}$ are to be considered. 

From  Fig.~\ref{tbh12}, it is found that for unstable gravitino with $ 30 \mbox{GeV} \lesssim m_{3/2} \lesssim 100  \mbox{GeV}$  (which is somewhat natural in the context of gauge hierarchy problem \cite{ellis1982flavour}) is allowed for the almost entire range of $M_{BH}$ ($10^{23} \mbox{GeV} \lesssim M_{BH} \lesssim  10^{32} \mbox {GeV} $). If unstable gravitino mass is below $30$ GeV (provided that LSP mass is also below that) then $M_{BH} > 10^{30}$ GeV is only possible. So depending on such lower unstable gravitino mass, using the constraint on $M_{BH}$, some allowed region in Figs.~\ref{short} (a)  and  ~\ref{1case2} (a) could be removed. However,  from Fig.~\ref{tbh12}, it is found that higher values of $m_{3/2} > 100 $ GeV are allowed only for relatively lower PBH mass.

\subsection{Stable gravitino}
Here, we discuss the constraints on gravitino (which is LSP) as dark matter which is non-thermally produced from PBHs. For dark matter produced from PBH, the constraints on the masses of dark matter with PBH mass have been earlier discussed in detail \cite{Fujita:2014hha}. Here, we are discussing gravitino, particularly as a dark matter for which the number density at different temperatures has been considered through Eq.~(\ref{gravpro}). Considering the production of gravitino from PBH and ignoring the production of gravitino from other supersymmetric particles, for a collisionless and non-thermally produced gravitino, one may write,
\[
\dot{n}_{3/2} + 3Hn_{3/2} = 0
\]
where the conservation of gravitino number in cosmic co-moving volume leads to
\[
n_{3/2} \propto a^{-3} \; .
\]
Using this one may write,
\be\label{to}
n_{3/2}(T_0) =\left(\frac{T_0}{T_{evap}}\right)^3 n_{3/2}(T_{evap})
\ee
where $n_{3/2}(T_0)$ is the present number density of the gravitino with the present temperature $ T_0 = 2.73 $ \text{K} $ \equiv 2.35 \times 10^{-13} \text{GeV}$.  Here, $n_{3/2}(T_{evap})$ is the number density of the gravitino at $T_{evap}$ which can be obtained from Eq.~(\ref{gravpro}) and $T_{evap}$ is obtained from Eq.~(\ref{tevap}). The present density parameter of the dark matter can be written as, 
\be\label{dark}
\Omega_{{DM},0} = \frac{m_{\text{3/2}} \, n_{3/2}(T_0)}{\rho_{{cri,0}}}  \simeq
\begin{cases}
\frac{m_{\text{3/2}} \,}{\rho_{{cri,0}}} \frac{n_{BH} \, 4 \pi \; {f_{3/2}} \; {M_{BH}}^2} { 3 \, M_{Pl}^2}
\left(\frac{T_0}{T_{evap}}\right)^3 & (m_{\text{3/2}} < T_{BH}) \\
\frac{m_{\text{3/2}} \,}{\rho_{{cri,0}}} \frac{n_{BH} f_{3/2}  \;{ M_{Pl}}^2}{48 \pi  \; {m_{3/2}}^2} \left(\frac{T_0}{T_{evap}}\right)^3 & (m_{\text{3/2}} \gtrsim T_{BH})
\end{cases}
\ee
where  critical density  $\rho_{{cri,0}} = 3.6 \times 10^{-47}  \text{GeV}^4 $. Using Eqs.~(\ref{gravpro}) and (\ref{tevap}) on the right-hand side of Eq.~(\ref{dark}), one can see that it varies on the values of both $M_{BH}$ and $m_{3/2}$. 

Depending on conditions on $T_{BH}$, $T_{evap}$ and $\Omega_{DM}$ as shown in Fig.~\ref{darkstable}, three regions with color shedding as grey, brown, and dark blue, have been shown. The present observed value of $\Omega_{{DM},0}$ has  upper bound as
$\Omega_{{DM},0} <0.25$. Satisfying this condition in Eq. ~(\ref{dark}), the three allowed regions have been plotted in Fig.~\ref{darkstable}.

However, the requirement of successful structure formation limits the velocity of dark matter particles, i.e. $\beta $ should be less than $4.9 \times 10^{-7}$  \cite{Viel2005}. Following \cite{Fujita:2014hha}, one may write for gravitino,

\begin{figure}[h]
\centering
\begin{tabular}{cc}
\includegraphics[scale=.7]{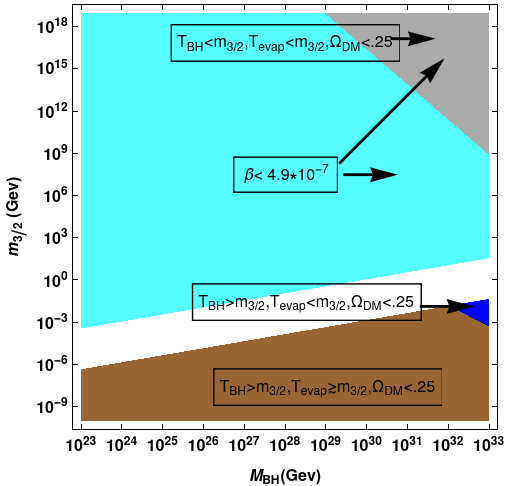}
\end{tabular}
\caption{(a)The allowed region of $m_{3/2}$ versus $M_{BH}$ is shaded in grey, satisfying both $\Omega_{{DM},0} <0.25$ and constraint on $\beta$  . The other three regions are shaded with brown, blue, and light blue respectively with their respective conditions as shown.}
\label{darkstable}
\end{figure}
 
\[
\beta \simeq \frac{T_0}{m_{3/2}\,T_{\text{evap}}} \times \frac{6 M_{\text{Pl}}^2}{8 \pi M_{BH}} < 4.9 \times 10^{-7}
\]
Using Eq.~(\ref{tevap}) for $T_{evap}$ above, one gets the condition on $\beta$ as the condition 
\begin{equation}\label{24}
\frac{M_{BH}^{1/2}}{m_{3/2}}<8.94\times 10^{14}
\end{equation}
For gravitino to be a viable candidate for dark matter, $\Omega_{DM,0}$ in Eq.~(\ref{dark}) is required to be $< 0.25$ as well as Eq.~(\ref{24}) corresponding to condition on $\beta$ is to be satisfied. The allowed region of $M_BH$ versus $m_{3/2}$ satisfying Eq.~(\ref{24}) is shown in a light blue shaded region. The region satisfying both Eqs.~(\ref{dark}) and (\ref{24}), is found in the upper grey shaded region which is finally the only allowed region of $M_{BH}$ and $m_{3/2}$.  
From Fig.~\ref{darkstable}, it is found that stable gravitinos of mass around the electroweak scale are not possible if it is produced from PBH.
So far we have discussed constraints on masses of gravitinos which are produced non-thermally from PBH. We like to make a short comment here for thermally produced gravitinos and PBH mass. 

 When $T_{evap} \gtrsim m_{3/2}$ then, the number density of thermally produced stable gravitino is  
 \begin{equation}
n_{3/2}^{therm}(T_{evap}) =  \frac{3 \, \zeta(3) \, f_{3/2} \,T_{evap}^3 }{4 \pi^2}   
 \end{equation}
 where $\zeta(3) \approx 1.2$ is Reimann zeta- function. Using this one may write,
\be\label{thermal}
n_{3/2}^{therm}(T_0) = \left(\frac{T_0}{T_{evap}}\right)^3 n_{3/2}^{therm}(T_{evap})
\ee
and
\be
\Omega_{{DM},0}^{therm} = \frac{m_{\text{3/2}} \, n_{3/2}^{therm}(T_0)}{\rho_{{cri,0}}} =\frac{m_{\text{3/2}} }{\rho_{{cri,0}}}   \frac{3 \, \zeta(3) \, f_{3/2} \,T_{evap}^3 }{4 \pi^2} \left(\frac{T_0}{T_{evap}}\right)^3 \lesssim 0.25 .
\ee
This implies $m_{3/2} \lesssim 6.8 \times 10^{-7}$ GeV. 
This is consistent with the requirement of  $T_{evap} \gtrsim m_{3/2}$ for allowed region of PBH mass.
Such gravitino mass will not play role in structure formation \cite{Viel2005}. 

With a higher reheating temperature of the universe which is $T_{evap}$ in our discussion, there will be too much overproduction of thermally produced gravitinos if $T_{evap}$ is higher than $10^{6-9}$ GeV. \cite{Nanopoulos1983, Krauss1983, Falomkin1984, Khlopov1984, Ellis1984, Juszkiewicz1985, Ellis1985b, Kawasaki1987, Khlopov1994, Moroi1993, Kawasaki1995, Bolz2001, Cyburt2003, Giudice1999, Kawasaki2005, Pradler2007, PhysLettB.648.224, ModPhysLettA.23.427, PhysRevD.78.065011, Cyburt2009}.   The abundance of gravitinos is related to $T_{evap}$ after inflation  \cite{PhysLettB.648.224,Bolz:2001a,Bolz:2001b,Moroi:1993}.
 For gravitinos as dark matter, depending on their mass, the upper bound of reheating temperature is obtained as $T_R \lesssim 10^7 \;{\mbox GeV}\; (m_{3/2}/0.1 \;{\mbox GeV})$. For $m_{3/2}$ lesser than 0.1 GeV, the upper bound of $T_R$ is lesser than $10^7$ GeV. If photino $\tilde{\gamma}$ and the gluino $\tilde{g}$ are typically lighter than the gravitino $\tilde{G}$, then  the decay processes $\tilde{G} \rightarrow \gamma + \tilde{\gamma}$ and $\tilde{G} \rightarrow g + \tilde{g}$ are possible. If the lifetime is long such that it decays during or after BBN, then the high energy photons emitted in gravitino decays could destroy light elements through photo-dissociation reactions and could increase $^4He$ abundance. All these processes put upper bounds on $T_R$ depending on gravitino mass. For $m_{3/2} \lesssim 100 $ GeV , $T_R \lesssim 10^{6-7} \mbox{GeV}$. For gravitino mass in the range $100 \, \text{GeV} \leq m_{3/2} \leq 1 \, \text{TeV}$, $T_R \lesssim 10^{7-9} \mbox{GeV}$ \cite{Chen:2007fv}.

However,  such overproduction of gravitinos (thermally produced) and also corresponding such higher $T_{evap}$ could be avoided naturally if PBH mass is higher than about $10^{25-27}$ GeV.  This naturally follows in case of soft leptogenesis with PBH which is possible only for Case \RN{2} and also shown in Fig.~\ref{mbhevap}.

\section{Conclusion}
\label{sec:summary}

If sneutrinos are produced from PBH evaporation then it is found that baryonic asymmetry could be obtained from sneutrino decays for soft SUSY breaking parameters $A$ and $B$ around the electroweak scale. We make a short comparative study below about soft leptogenesis with and without PBH. 

Without PBH, in soft leptogenesis, away from resonance, the stringent constraint on $A/M$ and $Y$ follows from  out- of-equilibrium condition in Eq.~(\ref{eq:out-of-equilibrium}). In this scenario, to get sufficient asymmetry without resonance (at resonance $B \sim \Gamma_\pm$) $M$  is required to be $\gtrsim 10^7$ GeV, otherwise from Eq.~(\ref{eq:out-of-equilibrium}), $A/M$ and $Y$ will be required to be even smaller than than $10^{-5}$ which will fail to produce sufficient $CP$ asymmetry shown in the Eq.~(\ref{eta}). However, for $M\gtrsim 10^7$, then there is problem from thermally produced gravitinos.

Without PBH, in soft leptogenesis, near resonance i.e. $B \sim \Gamma_\pm$, $M$ could be smaller than $10^7$ GeV and the lower value of $Y$ and $A/M$ may be compensated by resonance to get sufficient asymmetry in Eq.~(\ref{eta}). However, soft breaking $B$ parameter is required to be very small and fine-tuned with $\Gamma$.

In our work, with PBH, in soft leptogenesis, in case \RN{2} for which the asymmetry is obtained,  no out- of-equilibrium condition is required. Because of this $Y$ and $A/M$ is not constrained like soft leptogenesis without PBH. Soft breaking $A$ and $B$ parameter could be naturally around electroweak scale even for a lower values of $M \lesssim10^3$ GeV. Besides, for case II, leptonic asymmetry is obtained for PBH mass $M_{BH} \gtrsim 3.2 \times 10^{28}$ GeV or above as shown in Fig. \ref{short}(a) and \ref{1case2}(a). From Fig.~\ref{mbhevap}  of the paper, it is seen that $T_{evap}$ is happened to be less than $10^4$ GeV. So there is no problem due to thermally produced gravitinos in this scenario of soft leptogenesis. However, for gravitinos produced from PBH, the allowed region of PBH produced gravitino mass and corresponding PBH mass have been shown separately in  Fig.~\ref{tbh12} and \ref{darkstable}. So with PBH in soft leptogenesis, there is no requirement of resonance for $A$ and $B$ parameter around electroweak scale and also there is naturally no problem due to thermally produced gravitinos. Like case \RN{2}, for case \RN{3} also, the out-of-equilibrium condition is not required, as discussed in the context of Fig. \ref{allcase}. However, the expression of leptonic asymmetry for case \RN{3}, unlike case \RN{2}, is not given by Eq.~(\ref{full}) but by Eq.~(\ref{asymy}) which is suppressed by $M_{BH}^{5/2}$ and it is not possible to obtain the required baryonic asymmetry.

There are constraints on $Y$ versus $M$  based on collider searches on heavy leptons \cite{Abada:2022wvh} which has been taken into account in this work as shown in Fig.~\ref{short}(f) and Fig.~\ref{1case2}(f). Depending on the requirement of successful leptogenesis with sphaleron transition, with soft SUSY breaking parameters  around the electro-weak scale and such collider constraint, in  Fig.~\ref{short}(f), the lower bound of right handed neutrino mass $M \gtrsim 300$ GeV is obtained. In Fig.~\ref{clfvshort}, with sphaleron transition constraint, we have also discussed the MEG II experimental constraint on $|A|$ and $M$ values depending on values of  $m_{SUSY}$ - the typical mass scale of supersymmetric particles, due to non-observation of $\mu \rightarrow e \gamma$ decay.

Like sneutrino, gravitino also could be produced through PBH evaporation. Earlier authors \cite{Nanopoulos1983, Krauss1983, Falomkin1984, Khlopov1984, Ellis1984, Juszkiewicz1985, Ellis1985b, Kawasaki1987, Khlopov1994, Moroi1993, Kawasaki1995, Bolz2001, Cyburt2003, Giudice1999, Kawasaki2005, Pradler2007, PhysLettB.648.224, ModPhysLettA.23.427, PhysRevD.78.065011, Cyburt2009}  have discussed the mass bounds of gravitino (when it is thermally produced) depending on the reheating temperature. We have discussed the allowed region of PBH mass  $M_{BH}$ versus gravitino mass $m_{3/2}$ for unstable gravitino in detail. The allowed region for stable gravitino mass $m_{3/2}$ versus $M_{BH}$ is almost similar to the allowed region shown earlier in \cite{Fujita:2014hha} for PBH-produced dark matter.  It is found in Fig. \ref{tbh12} that for almost the entire range of $M_{BH}$ $10^{23}-10^{32}$, the unstable gravitino mass $ 30  \mbox{GeV} \lesssim m_{3/2} < 100 $ GeV is allowed.
However, if unstable gravitino mass is below $30$ GeV (provided that some LSP mass is also below that)  then $M_{BH} > 10^{30}$ GeV is only possible.  From Fig. \ref{tbh12} and \ref{darkstable}, it is found that if gravitinos are produced from PBH evaporation, unstable gravitino mass around the electroweak scale, is allowed for almost the entire range of PBH mass, while stable gravitino with mass around electroweak scale, is  not possible.

\acknowledgments
SK thanks the Council of Scientific and Industrial Research (CSIR), India for financial support through
Senior Research Fellowship (Grant No. 09/466(0209)/2018-EMR-I). SK also thanks Imtiyaz Ahmed Bhat, Yogesh, and Kunal Pandey for their helpful discussions.

\bibliography{susy}
\bibliographystyle{apsrev4-1}
\end{document}